\documentclass[onecolumn,preprint,aps,superscriptaddress,nofootinbib]{revtex4}
\usepackage{graphicx,amsmath}
\usepackage{hyperref}
\usepackage{amssymb}
\usepackage[usenames, dvipsnames]{color}
\usepackage[normalem]{ulem}
\usepackage{multirow}
\usepackage{siunitx}

\newcommand{\beq}{\begin{equation}}
\newcommand{\eeq}{\end{equation}}
\newcommand{\bea}{\begin{eqnarray}}
\newcommand{\eea}{\end{eqnarray}}

\newcommand{\ev}{\mathrm{eV}}
\newcommand{\kev}{\mathrm{keV}}

\newcommand{\tot}{\mathrm{tot}}
\newcommand{\mpc}{\mathrm{Mpc}}
\newcommand{\mH}{\mathcal{H}}

\newcommand{\Pl}{\mathrm{Pl}}
\newcommand{\SM}{\mathrm{SM}}

\newcommand{\eq}{\mathrm{eq}}
\newcommand{\cheq}{\mathrm{ch}}

\newcommand{\cdm}{\mathrm{cdm}}
\newcommand{\cc}{\mathrm{can}}
\newcommand{\nr}{\mathrm{nr}}
\newcommand{\eff}{\mathrm{eff}}

\newcommand{\Sec}[1]{Sec.~\ref{#1}}

\newcommand{\App}[1]{Appendix~\ref{#1}}

\newcommand{\Fig}[1]{Fig.~\ref{#1}}

\newcommand{\Eq}[1]{Eq.~(\ref{#1})}
\newcommand{\Eqs}[2]{Eqs.~(\ref{#1}) and (\ref{#2})}
\newcommand{\Eqst}[2]{Eqs.~(\ref{#1})-(\ref{#2})}
\newcommand{\eg}{\textit{e.g.}\ }
\newcommand{\ie}{\textit{i.e.}\ }

\newcommand{\threeto}{\ensuremath{3\rightarrow 2\ }}
\newcommand{\twoto}{\ensuremath{2\rightarrow 2\ }}
\def\f0{\ensuremath{f^{(0)}}}

\newcommand{\bl}{\left}
\newcommand{\br}{\right}

\newcommand{\ignore}[1]{}

\makeatletter
\def\gsim{\mathrel{\lower2.5pt\vbox{\lineskip=0pt\baselineskip=0pt
           \hbox{$>$}\hbox{$\sim$}}}}
\def\lsim{\mathrel{\lower2.5pt\vbox{\lineskip=0pt\baselineskip=0pt
           \hbox{$<$}\hbox{$\sim$}}}}
\makeatother

\begin{document}
\setlength{\unitlength}{1mm}

\title{Cannibal Dark Matter and Large Scale Structure}

\author{Manuel A. Buen-Abad}
 \email{buenabad@bu.edu}
\affiliation{Physics Department, Boston University, Boston, MA 02215, USA}
\author{Razieh Emami}
 \email{iasraziehm@ust.hk}
\affiliation{Hong Kong University of Science and Technology, Clear Water Bay, Kowloon, 999077 Hong Kong, China}
\author{Martin Schmaltz}
\email{schmaltz@bu.edu}
\affiliation{Physics Department, Boston University, Boston, MA 02215, USA}

\vskip.4in

\begin{abstract}

Cannibals are dark matter particles with a scattering process that allows three particles to annihilate to two. This exothermic process keeps the gas of the remaining particles warm long after they become non-relativistic. A cannibalizing dark sector which is decoupled from the Standard Model naturally arises from a pure-glue confining hidden sector. It has an effective field theory description with a single massive interacting real scalar field, the lightest glueball. Since warm dark matter strongly suppresses growth of structure cannibals cannot be all of the dark matter. 
Thus we propose a scenario where most dark matter is non-interacting and cold but about 1 percent is cannibalistic. We review the cannibals' unusual scaling of the temperature and energy and number densities with redshift and generalize the equations for the growth of matter density perturbations to the case of cannibals. We solve the equations numerically to predict the scaling of the Hubble parameter and the characteristic shape of the linear matter power spectrum as a function of model parameters. Our results may have implications for the $\sigma_8$ and $H_0$ problems.

\end{abstract}

\maketitle

\section{Introduction}
\label{sec:intro}

Dark matter could be a single species of particles with only gravitational interactions as in the cosmological standard model, $\Lambda$CDM. Alternatively, it might have multiple components. If there is a dominant non-interacting component then other components can have interesting non-gravitational interactions. Recent observations of the Cosmic Microwave Background (CMB) and matter power spectrum (MPS) are already sensitive to non-standard dark matter components which comprise only a few \% of the total, stage 4 experiments will be able to push the sensitivity below the percent level. Interestingly, precision fits with current cosmological data show some tension with predictions of $\Lambda$CDM for the expansion rate of the universe $H_0$ \cite{Riess:2016jrr,Bonvin:2016crt} and the amplitude of fluctuations in the MPS on galaxy cluster scales, $\sigma_8$ \cite{Heymans:2013fya,Joudaki:2016mvz,Ade:2015fva,Ade:2013lmv,Kohlinger:2017sxk,Joudaki:2017zdt}\footnote{For recent work motivated by these discrepancies see \cite{Buen-Abad:2015ova,Lesgourgues:2015wza,Chacko:2016kgg,Poulin:2016nat,MacCrann:2014wfa,Canac:2016smv,Bernal:2016gxb,Chudaykin:2016yfk,Archidiacono:2016kkh,Joudaki:2016kym,Buen-Abad:2017gxg,Raveri:2017jto,Lancaster:2017ksf,Oldengott:2017fhy,Ko:2016fcd,Ko:2016uft,Ko:2017uyb,Chacko:2018vss,Poulin:2018zxs,Pan:2018zha}.}. Motivated by the significant projected improvement in measurements of the MPS we propose and explore the possibility that a small component of dark matter is ``cannibalistic''.

Cannibal dark matter consists of massive particles with an efficient number-changing self-interaction \cite{Carlson:1992fn}. The most important process that such interactions mediate is from three particles in the initial state to two particles in the final state. In such a \threeto process mass is turned into kinetic energy of the outgoing particles which heats the gas of particles\footnote{Cannibals cannot constitute the entirety of the dark matter in the Universe precisely because they are heated up by their self-interactions and that interferes with the formation of structure \cite{Machacek:1994vg,deLaix:1995vi}. Proposed solutions to this problem are to let cannibalism end much before matter domination \cite{Soni:2016gzf,Pappadopulo:2016pkp,Dey:2016qgf,Bernal:2015ova} or to cool it through couplings to the Standard Model, like in the ELDER \cite{Kuflik:2015isi} or SIMP \cite{Hochberg:2014dra} paradigms. The SIMP mechanism has been the object of intense study in recent years \cite{Hochberg:2014kqa,Bernal:2015bla,Bernal:2015xba,Bernal:2017mqb,Lee:2015gsa,Kamada:2016ois,Choi:2017mkk,Choi:2017zww,Choi:2018iit}.}. If there are also rapid \twoto interactions the cannibalizing particle gas remains in thermal and chemical equilibrium, and can be described by the Boltzmann distribution with a temperature $T(a)$ and vanishing chemical potential. Because of the cannibalization process the temperature drops only logarithmically with the scale factor $T/m \sim 1/\log a$. This is very different from the case of non-relativistic matter which cools very quickly, $T/m \sim 1/a^2$. Cannibal matter also has an unusual scaling of its number and energy densities. The number density dilutes like $n_\cc\sim 1/(a^3 \log a)$ where the $1/a^3$ is the usual volume dilution while the $1/\log a$ comes from the cannibalization. Ignoring kinetic energy, the energy density is then simply $\rho_\cc \approx m\, n_\cc$. Thus the energy density of cannibals scales intermediate between  ordinary matter for which $\rho_{m} \sim 1/a^3$ and radiation where $\rho_{r} \sim 1/a^4$. Note that for these scalings to hold it is necessary that the cannibal particles are isolated from all other sectors, \ie no significant interactions, so that any heat produced from cannibalization does not dissipate to other sectors.

We now discuss the impact of the cannibal fluid on cosmology with particular attention to the MPS. First, note that since the cannibal temperature decays very slowly the cannibal fluid has significant pressure $P/\rho \approx T/m$. This pressure prevents growth of density perturbations in the cannibal fluid, instead one obtains ``cannibal acoustic oscillations". Overdensities in the cannibal fluid remain small and make only negligible contributions to the gravitational potential.  On the other hand, the cannibal fluid does contribute to the overall energy density of the universe which determines the Hubble expansion rate. Since the gravitational potential drives the growth of structure whereas the Hubble expansion acts to slow it (``Hubble friction") the net effect of the cannibal fluid is to suppress the MPS. This is the main result of our paper. 

In Section \ref{sec:cosmo} we derive this result quantitatively. The connection to the physical explanation in the previous paragraph will become clear after we derive the M\'{e}sz\'{a}ros equation for the growth of cold dark matter (CDM) perturbations $\delta_\cdm$ in the presence of the cannibal fluid:
\beq\label{eq:cdmperts}
	a^2 \delta_\cdm'' + \frac{3}{2} a \delta_\cdm' - \frac{3}{2} \frac{\rho_\cdm}{\rho_\cdm+\rho_\cc} \delta_\cdm = 0 \ .
\eeq

This equation is valid during matter domination and for perturbations which are deep inside the horizon.%
\footnote{We have simplified further by dropping terms which are suppressed by $T/m$ of the cannibals.}  
Here the derivatives are with respect to the scale factor $a$, and $\rho_\cdm$ and $\rho_\cc$ are the background (average) energy densities of the cold dark matter and the cannibals, respectively. For zero cannibal energy density this has the usual linear growth of the matter perturbations $\delta_\cdm \sim a$ as a solution. Expanding for small energy density in cannibals $\rho_\cc \ll \rho_\cdm$ one finds a suppressed rate of growth: $\delta_\cdm \sim a^{1-\gamma}$ with $\gamma = \frac35\, \rho_\cc/ \rho_\cdm$. Given that current data suggest a suppression of matter perturbations by $\sim 5\%$ and that the universe expands by a factor of $a_{\rm today}/a_{\rm equality} \sim 10^3$ during matter domination we see that the preferred parameter space should have on the order of 1\% of matter in cannibals, \ie a fraction $\rho_\cc/ \rho_\cdm \sim 1\%$ which slowly changes in time due to the extra $1/\log a$ in $\rho_\cc$.

The minimal field theoretic model which exhibits cannibalism has a real scalar field with the Lagrangian
\bea\label{eq:lagrange}
	\mathcal{L}=\frac12 (\partial \phi)^2 - \frac12 m^2 \phi^2 - \kappa_3 m \lambda\frac{\phi^3}{3!} - \kappa_4 \lambda^2 \frac{\phi^4}{4!}  \ .
\eea

In this minimal cannibal (MC) model $m$ is the mass of the particle, $\lambda$ denotes the overall strength of $\phi$-interactions and $\kappa_{3,4}$ are numbers which we will take to be of order 1. The interactions mediate $\phi$-number preserving $\phi\phi \rightarrow \phi \phi$ processes as well as $\phi$-number changing processes such as $\phi\phi\phi \rightarrow \phi\phi$ (with a rate proportional to $\lambda^6$). At temperatures above the $\phi$ mass the $\phi$ particles can be described by an interacting relativistic fluid in equilibrium. Once the $\phi$-fluid cools below the mass of the particles the \threeto cannibalism interaction starts processing mass into temperature. This slows the cooling of the fluid.  The fluid remains in thermal equilibrium during cannibalization because the \twoto interactions are very rapid compared with the cannibal interactions and with the expansion rate of the universe and rethermalize the fluid. Furthermore, since the $\phi$ particles are isolated from all other fluids (such as the Standard Model and the cold dark matter) and heat cannot be dissipated to the other sectors the comoving entropy in the $\phi$-fluid is conserved. Eventually, at late times, the number density of $\phi$ particles becomes too small for the \threeto interactions to compete with the expansion rate and they turn off, bringing cannibalism to an end. At that point the surviving particles become cold dark matter, their number density diluting with the volume and their temperature dropping rapidly proportional to $1/a^2$.

This thermal history is summarized in the following table: the $\phi$-fluid cools like radiation while its temperature is above the $\phi$ mass,  at $a\sim a_\cc$ it enters the cannibalistic phase where the temperature drops logarithmically, and at $a\sim a_\nr$ the \threeto interactions decouple and it cools like ordinary non-relativistic matter.
\bea
\begin{array}{c|c|c}\label{tab:scaling}
{\rm relativistic} & {\rm cannibal} & {\rm non\!-\!relativistic} \\
a<a_\cc & a_\cc<a<a_\nr & a_\nr<a \\
\hline
T\sim  1/a & T\sim {1}/{\log a } & T\sim 1/a^2 \\
\rho \sim  1/a^4 & \rho \sim {1}/(a^3 \log a) & \rho \sim 1/a^3
\end{array} \nonumber
\eea

\begin{figure}[!htbp]%
    \centering
    \includegraphics[width=0.55\textwidth]{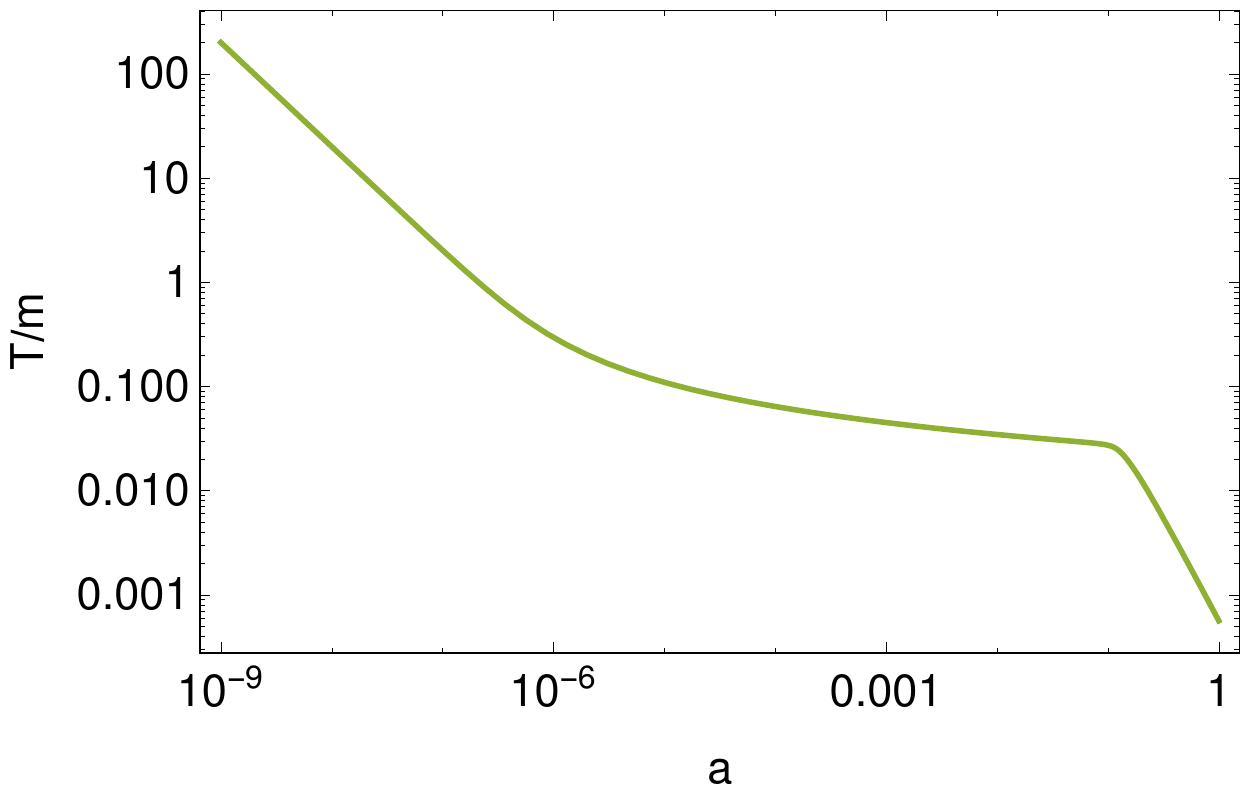}
    \caption{Temperature to mass ratio as a function of scale factor $a$ for the minimal cannibal (MC) model. The temperature drops like $1/a$ while the particles are relativistic, it drops logarithmically in $a$ while the particles cannibalize, and it drops like $1/a^2$ after the cannibalizing interaction decouples and the particles cool like ordinary non-relativistic matter. The temperature curve shown here was found by solving the background equations (\ref{eq:A32}) numerically and includes the decoupling of \threeto interactions.}%
    \label{fig:temp}
\end{figure}

In \Fig{fig:temp} we plot the temperature-to-mass ratio as a function of scale factor for an example point in parameter space of the minimal cannibal model. Note the transition from relativistic behavior to cannibalism at $T/m\sim1/3 \  \leftrightarrow \ a_\cc \sim 10^{-6}$ and the decoupling transition to non-relativistic matter at $a_\nr \sim 10^{-1}$.  The ratio of scale factors between start and end of the cannibalistic phase $a_\nr/a_\cc\sim 10^5$ depends on the strength of the interaction $\lambda$. We will be interested in models where $\lambda$ is strong (between 1 and $4\pi$); then the duration of cannibalism $a_\nr/a_\cc$ is between $10^{-4}$ and $10^{-5}$ with only a mild dependence on other model parameters.  

\begin{figure}[!htbp]%
    \centering
    \includegraphics[width=0.75\textwidth]{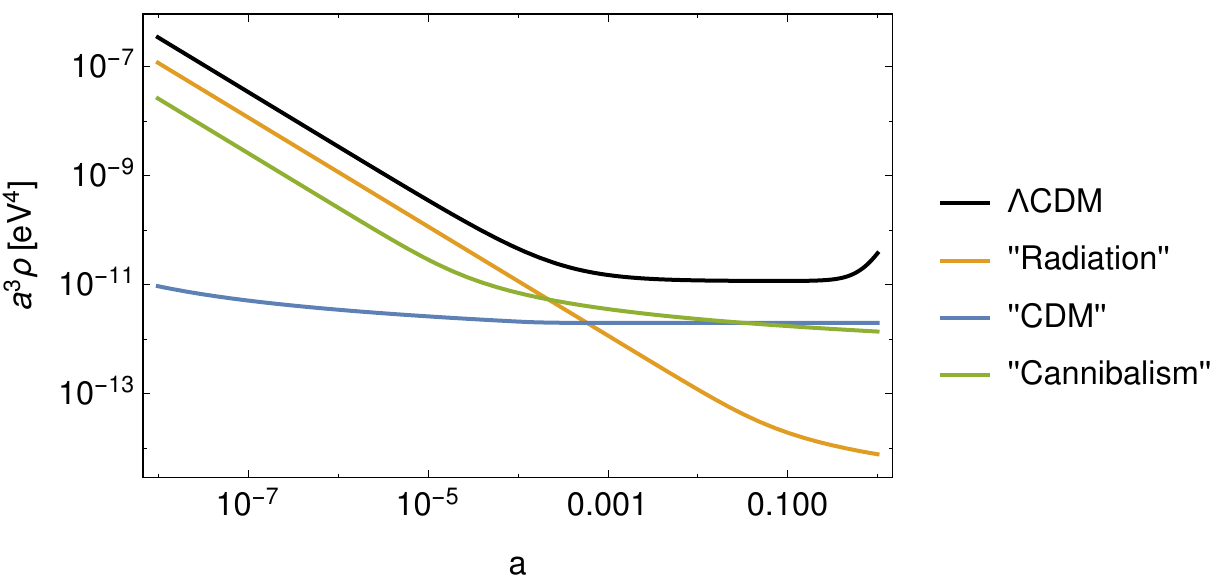}
    \caption{Energy densities for MC models with mass and temperature chosen such that $\rho_\cc < \rho_{\Lambda\mathrm{CDM}}$. A MC model for which cannibalism occurs throughout matter domination is shown in green with its characteristic $\rho_\cc \sim 1/(a^3 \log a)$ dilution. The orange model has a late onset of cannibalism, making the $\phi$-fluid behave like radiation throughout most of the history of the Universe. In the blue model the cannibalism phase is shifted very early so that cannibalism stops before matter domination. Then the $\phi$-fluid behaves like cold dark matter. For comparison, we also show the total energy density in the components of $\Lambda$CDM (black).}
    \label{fig:3rhos}
\end{figure}

From  preceding discussions it is clear that we can choose parameters in the cannibal sector such that the cannibalistic phase overlaps with the matter-dominated era of the universe. This choice of parameters is the most interesting because then the cannibals suppress the matter power spectrum. We dedicate most of this paper to its study. 
In \Fig{fig:3rhos} we show the evolution of the energy density of the cannibal fluid (green) in a model where the cannibal transition happens at $a_c \sim 10^{-5}$ and decoupling at $a_\nr \sim 1$. For comparison we show the total energy density in the $\Lambda$CDM components (black) with its radiation-, then matter-, and finally cosmological constant-dominated scale dependence. We also show the energy densities for two different MC models: one where the cannibal transition happens well after matter-radiation equality (orange) so that the cannibals act as radiation while they have significant energy densities. Such a model is indistinguishable from a model with extra neutrinos $\Delta N_\eff$. The other model (blue) is one in which the cannibal transition happens so early that the cannibal interactions already decouple before matter-radiation equality. Then the cannibals behave like ordinary cold dark matter.

The MC model in \Eq{eq:lagrange} is ugly because the cannibal mass is unprotected from quadratically divergent quantum corrections and has a naturalness problem. Fortunately, natural UV completions are easy to construct. Our favorite is a simple non-Abelian gauge sector without matter (\ie pure-glue). Such a model has a single coupling constant, the gauge coupling. The theory is asymptotically free in the UV. The gauge coupling becomes strong in the IR, the theory confines and the spectrum is one of glueball resonances. The effective low-energy description below the confinement scale is the MC model \Eq{eq:lagrange} where $\phi$ is the lightest glueball, $m$ is its mass, and $\lambda \sim 4\pi$. In addition to the renormalizable interactions shown in \Eq{eq:lagrange} one also obtains higher-dimensional couplings of the form $\lambda^{n-2} \phi^n/m^{n-4}$ which contribute to scattering with the same parametrics as the renormalizable couplings. The cannibalism phase is not sensitive to the precise form of the interactions: what matters is that the number-changing transitions are faster than the Hubble expansion. Then the cannibal fluid satisfies thermal and chemical equilibrium and its evolution becomes independent of the details of the spectrum of glueballs and interactions. Note also that this UV completion very naturally explains the absence of couplings between $\phi$ and the Standard Model. In the UV theory gauge invariance forbids any renormalizable coupling between the two sectors. We describe such UV completions and study the dependence of our results on the UV completion in Section \ref{sec:simplest}.

Finally, we do not consider but cannot resit mentioning the possibility that the cold dark matter required in our model might be ``Baryons" or ``Mesons" made of heavy dark quarks charged under the dark gauge group \cite{Boddy:2014qxa} although important details of the confining phase transition and entry into the cannibal phase would change from what we study in this paper. 

We study the MC model of \Eq{eq:lagrange} and its thermal history in \Sec{sec:thermo} where we also estimate the boundaries of the preferred parameter space. Within these boundaries we compute the effects of cannibalism on the matter power spectrum  in \Sec{sec:cosmo}. \Sec{sec:simplest} gives possible UV realizations of the MC model in terms of simple confining (pure-glue) non-Abelian gauge theories. We also study the depedence of our results on the UV completion of the MC model. In the Conclusions (\Sec{sec:conc}) we discuss the shape of the predicted MPS as a function of model parameters. We review the derivation of the background and perturbation equations for the cannibal fluid starting from the Boltzmann equation in an Appendix; our results agree with those given in \cite{Ma:1995ey}.

\section{The minimal cannibal: thermal history and parameters}
\label{sec:thermo}

In this Section we study the thermal history of the MC model fluid, identify the most useful parameters to describe it, and explore their parameter space. In order to do this we need to consider what the properties of the cannibal fluid are.

During its relativistic and cannibalistic phases the $\phi$-fluid is in both thermal and chemical equilibrium. This means that its phase space distribution function $f(p,a)$ is entirely parameterized by the mass of the particles $m$ and the temperature $T$ of the fluid\footnote{Here $T(a)$ denotes the temperature of the cannibal fluid which may be different from the temperature of the Standard Model (e.g. photons).}:
\beq
	f(p,a)=\frac{1}{e^{E/T(a)}-1} \ ,
	\label{eq:distribution}
\eeq
where $E=\sqrt{m^2 + p^2}$ is the energy of the $\phi$ particles. Here we only consider the homogeneous and isotropic background of the cannibal fluid which means that $f$ does not depend on position. We will study $x$-dependent perturbations about this background in the following Section.  The time dependence of $f$, encoded in the scale factor $a(t)$, arises solely from that of the temperature. All other background quantities that describe the $\phi$-fluid (such as energy and number densities) are momentum integrals of $f$, and therefore they depend on the two parameters $m$ and $T(a)$. 

Since the cannibal fluid has no interactions with other fluids its (comoving) entropy $S_\cc$ is conserved. This makes $S_\cc$ a useful parameter of the MC model. We now derive formulae for the temperature and energy density of the $\phi$-fluid in terms of the model parameters $m$ and $S_\cc$. From the Second Law of Thermodynamics:
\beq\label{eq:entropy_def}
	S_\cc = a^3 \frac{\rho_\cc + P_\cc}{T} \ .
\eeq

In the relativistic limit, $T\gg m$, the phase space distribution function, \Eq{eq:distribution}, is easily integrated to obtain expressions for the energy density $\rho = \frac{\pi^2}{30} T^4$ and pressure $P = \rho/3$ so that:
\beq\label{eq:entropy_uv}
	S_\cc = a^3 \frac{2 \pi^2}{45} T^3 \ .
\eeq
Solving for $T$ we find:
\beq
	T = \bl( \frac{45}{2 \pi^2} \br)^{1/3} \frac{S_\cc^{1/3}}{a} \ , \quad
	\rho_\cc = \frac{3}{4} \bl( \frac{45}{2 \pi^2} \br)^{1/3} \frac{S_\cc^{4/3}}{a^4} \ . \label{eq:temp_rho_uv}
\eeq
Note that $T \sim 1/a$ and $\rho_\cc \sim 1/a^4$, as expected for radiation components.\footnote{\Eq{eq:entropy_uv} contains a factor of $g$ that accounts for the degrees of freedom of the dark sector. This factor is 1 in the $\phi$ cannibal model but will be different in UV completions.}

Once $T \sim m$ the $\phi$-fluid enters its cannibalistic phase. After the temperature drops sufficiently far below the mass an expansion in $T/m$ becomes appropriate, and the dominant contribution to the energy density comes from the mass of the particles, $\rho_\cc \approx m n_\cc$, where $n_\cc = (\frac{mT}{2 \pi})^{3/2} e^{-m/T}$ is the equilibrium number density of $\phi$. The contribution of the pressure $P_\cc \approx T n_\cc$  to the entropy in \Eq{eq:entropy_def} is smaller by $T/m$ relative to $\rho_\cc$ so that:
\bea\label{eq:entropy_ir}
	S_\cc \simeq a^3 \frac{\rho_\cc}{T} \simeq \frac{a^3 m^3}{(2 \pi)^{3/2}} \bl( \frac{T}{m} \br)^{1/2} e^{-m/T} \ ,
\eea
and solving for $T$ and $\rho_\cc$ in a leading-log approximation we have:
\beq
	T \simeq \frac{m}{3 \log \bl( \frac{m \, S_\cc^{-1/3}}{\sqrt{2 \pi}} \, a \br)} \ , \quad
	\rho_\cc \simeq \frac{m \, S_\cc}{3 a^3 \log\bl( \frac{m \, S_\cc^{-1/3}}{\sqrt{2 \pi}} \, a \br)} \ . \label{eq:temp_rho_ir}
\eeq
Note that $T\sim 1/\log a$ and $\rho_\cc \sim 1/(a^3 \log a)$ as stated in the previous Section.

Having written $T$ and $\rho_\cc$ as functions of $a$ and the parameters $m$ and $S_\cc$, we now study the parameter space. Our goal is to estimate the values of the parameters for which the cannibal sector suppresses the matter power spectrum by about the amount that is preferred by the $\sigma_8$ measurements. As shown in the Introduction this requires a fraction of dark matter energy density in the $\phi$-fluid $f_\cc \equiv \rho_\cc/\rho_\cdm \sim \mathcal{O}(1\%)$. Of course, since $\rho_\cdm \sim 1/a^3$ but $\rho_\cc \sim 1/(a^3 \log a)$, this fraction evolves as $f_\cc \sim 1/\log a$. But the change in $f_\cc$ during matter domination is small enough (of order of a few) that we ignore it for the purpose of estimating the rough region of $m$ - $S_\cc$ parameter space where we can expect to find good fits. The good region of parameter space is the one in which the cannibalism phase overlaps with matter domination, which corresponds to conditions on $a_c$ and $a_\nr$, and in which $f_\cc \sim \mathcal{O}(1\%)$. In the remainder of this Section we use these conditions to derive that
\bea\label{eq:parameter_bounds}
 \ev \lsim m \lsim  \kev\, , \quad \frac{S_\cc}{S_\SM} \sim\,  0.1\, \bl[\frac{1\ev}{m}\br]\ .
\eea
A reader who is not interested in the following somewhat tedious derivation of these boundaries of the relevant parameter space may skip ahead to \Sec{sec:cosmo} where we derive and solve the density perturbation equations. 

We first derive the lower bound on $m$. Define the scale factor $a=a_\cc$ at which $T(a_\cc) \equiv m/3$, \ie where the $\phi$-fluid stops being relativistic and starts cannibalizing. From \Eq{eq:temp_rho_ir} we obtain $a_\cc \sim 10\, S_\cc^{1/3}/m$. Since we want cannibalism to act during matter domination, we require the start of cannibalism to be before matter-radiation equality, \ie $a_\cc < a_\eq$. Ignoring the $\log a$ dependence (for simplicity) and using $a_\cc \sim 10\, S_\cc^{1/3}/m$ we express $\rho_\cc$ in \Eq{eq:temp_rho_ir} in terms of $a_\cc$
\beq\label{eq:rho_ac}
	\rho_\cc \sim \frac{m^4}{10^3 (a/a_\cc)^3} \ .
\eeq
We solve this for $m$, substitute $\rho_\cc = f_\cc \rho_\cdm$, evaluate it today ($a=1$) and impose $a_\cc < a_\eq$ to obtain:
\beq\label{eq:mlower1}
	m^4 \sim 10^3 \times \frac{f_{\cc,\, 0} \, \rho_{\cdm,\, 0}}{a_\cc^3} > 10^3 \times \frac{f_{\cc,\, 0} \, \rho_{\cdm,\, 0}}{a_\eq^3} \ ,
\eeq
which for $a_\eq \approx 3\times 10^{-4}$ gives the lower bound:
\beq\label{eq:mlower2}
	m \gsim \,1\, \ev \times \bl[ \frac{f_{\cc,\, 0}}{0.01} \br]^{1/4} \bl[ \frac{\rho_{\cdm,\, 0}}{10^{-11}\, \ev^4} \br]^{1/4} \ .
\eeq

At the edge of the preferred parameter space, when $m$ saturates the bound, the $\phi$-fluid enters its cannibalistic phase right at matter-radiation equality. Then the UV completion of the $\phi$ model is needed to determine the cannibal sector energy density for $a < a_\eq$. Thus in this case the matter power spectrum is sensitive to details of the UV completion such as the glueball spectrum and the size of the UV gauge group.  We will study this model dependence in \Sec{sec:simplest}. For masses much smaller than the bound the cannibal sector is still relativistic at $a_\eq$. In that case the cannibal fluid behaves like extra radiation ($\Delta N_\eff$) at the time of the CMB. Imposing observational bounds on $\Delta N_\eff$ bounds the energy density in the cannibal fluid at $a_\eq$ and by the time cannibalism turns on at
$a_\cc > a_\eq$ the energy density in the cannibal fluid has already become negligible compared with that in $\Lambda$CDM (orange curve of \Fig{fig:3rhos}). Thus this is not a region in parameter space that we are interested in. 

We can also derive an upper bound on $m$. To do so we first solve for the scale factor $a_\nr$ when the \threeto interactions decouple and the $\phi$-fluid transitions from cannibal behavior to standard non-relativistic behavior. Dimensional analysis allows us to estimate the non-relativistic \twoto and \threeto scattering cross sections in the $\phi$ theory from \Eq{eq:lagrange}:
\bea\label{eq:scatter}
	\sigma_{22} v \approx \frac{\alpha^2}{m^2} & \Rightarrow & \Gamma_{22} \equiv n_\cc \langle \sigma_{22} v \rangle \approx \frac{\alpha^2}{m^{3}}\, \rho_\cc \ , \\
	\ \sigma_{32} v^2 \approx \frac{\alpha^3}{m^5} & \Rightarrow & \Gamma_{32} \equiv n_\cc^2 \langle \sigma_{32} v^2 \rangle \approx \frac{\alpha^3}{m^{7}}\, \rho_\cc^2 \ ;\label{eq:scatter32}
\eea
where $\alpha \sim \lambda^2/(4\pi)$, $\Gamma_{ij}$ are the $i \rightarrow j$ interaction rates, and we have been cavalier with factors of order 1 and $\pi$. Keeping in mind a strongly coupled UV completion of the cannibal sector we expect $\alpha$ somewhere between 1 and $4\pi$.

Eventually $\Gamma_{32}$ cannot keep up with the rate of expansion of the Universe $H$ and the \threeto interactions decouple and  cannibalism stops at $a_\nr$. Setting $\Gamma_{32} = H$ and using \Eqs{eq:scatter32}{eq:rho_ac} we can solve for the duration of the cannibalistic phase
\bea\label{eq:duration1}
	\frac{a_\nr}{a_\cc} \approx \frac{\alpha^{1/2}}{10} \bl( \frac{m}{H(a_\nr)} \br)^{1/6} \ .
\eea

Note the small exponent of 1/6. This shows that the duration of the cannibalism phase is only weakly dependent on the model parameters $m$ and $S_c$. In particular, the duration of the cannibalism phase is rather insensitive to when the decoupling occurs. For example, if cannibalism ends at matter-radiation equality ($a_\nr = a_\eq$) then $(H(a_\eq)/\ev)^{1/6} \sim 10^{-5}$; whereas if it ends today ($a_\nr = 1$), then $(H_0/\ev)^{1/6} \sim 10^{-6}$; a change of only one order of magnitude. The duration of the cannibalistic phase is therefore between 4 and 5 decades in the scale factor:
\beq\label{eq:duration2}
	\frac{a_\nr}{a_\cc} \approx 10^5 \times \bl[ \frac{\lambda}{4 \pi} \br] \bl[ \frac{m}{1\ev} \br]^{1/6} \bl[ \frac{10^{-33}\, \ev}{H(a_\nr)} \br]^{1/6} \ .
\eeq

We will use the approximation $a_\cc \sim 10^{-5} a_\nr$. Substituting this in \Eq{eq:rho_ac} yields:
\beq\label{eq:mupper1}
	m^4 \sim 10^{18} \times \frac{f_{\cc,\, 0} \, \rho_{\cdm,\, 0}}{a_\nr^3} \ .
\eeq

In order to find an upper bound on the interesting range of $m$ we impose a condition on $a_\nr$, the scale factor when cannibalism stops. Demanding that cannibalism lasts throughout matter domination and does not end before today so as to maximize the suppression of the MPS is a possibility.  But this is really too aggressive because even when cannibalism stops midway through matter domination the MPS is suppressed relative to $\Lambda$CDM. We impose - admittedly somewhat arbitrarily - that $a_\nr\gsim 10^{-2}$. This together with \Eq{eq:mupper1} implies
\beq\label{eq:mupper3}
	m \lsim \,1\, \kev \times \bl[ \frac{f_{\cc,\, 0}}{0.01} \br]^{1/4} \bl[ \frac{\rho_{\cdm,\, 0}}{10^{-11}\, \ev^4} \br]^{1/4} \ .
\eeq

For masses much larger than this bound the end of cannibalism occurs too close to (or before) matter-radiation equality, so that the $\phi$-fluid clusters like cold dark matter during matter domination as discussed in the previous section (blue curve in \Fig{fig:3rhos}).
Comparing \Eq{eq:mupper3} and \Eq{eq:mlower2} we see the range of masses, $\ev < m < \kev$, which satisfies both constraints.

Having restricted the mass of the $\phi$ particles to a range for which cannibalization has an interesting effect on the MPS we now focus our attention on the other parameter of the MC model, the entropy. Starting again from the relationship between the energy density and the entropy in  \Eq{eq:temp_rho_ir}, approximating $\log a_\cc^{-1}\sim 8$, demanding that the energy density in cannibals be a small fraction $f$ of that in the the $\Lambda$CDM sector, and evaluating energy densities today we obtain 

\beq\label{eq:entropyofm}
	S_\cc \sim \frac{S_\SM}{10} \bl[ \frac{2.2\times 10^{-11}\, \ev^3}{S_\SM} \br] \bl[ \frac{f_{\cc,\, 0}}{0.01} \br] \bl[ \frac{\rho_{\cdm,\, 0}}{10^{-11} \, \ev^4} \br] \bl[ \frac{1\, \ev}{m} \br] \ ,
\eeq
where we have chosen to write the comoving cannibal sector entropy $S_\cc$ in terms of the comoving entropy in the Standard Model sector today, $S_\SM =2.2 \times 10^{-11} \ev^3$. One sees that the values of $S_\cc$ which give the correct suppression of the MPS are
inversely proportional to $m$.

Finally, let us verify that thermal (kinetic) equilibrium is maintained until today in the region of parameter space we have obtained. We must check that the rate of \twoto interactions is faster than the expansion rate of the Universe. From \Eq{eq:scatter}
\beq
	\Gamma_{22, 0} \approx 10^{22} \bl[ 10^{-33} \, \ev \br] \bl[ \frac{\alpha}{4\pi}\br]^2 \bl[ \frac{1 \, \ev}{m} \br]^{3} \bl[ \frac{f_{\cc,0}}{0.01} \br] \bl[ \frac{\rho_{\cdm,0}}{10^{-11} \, \ev^4} \br] \ ,
\eeq
clearly bigger than $H_0 \sim 10^{-33} \ \ev$. This is not surprising because \twoto interactions are much more rapid than \threeto interactions which are suppressed by an additional power of the particle number density.

In summary, in order for the cannibalistic phase to overlap with matter domination and suppress the matter perturbations at galaxy cluster scales by about 5\%  we need $f_{\cc,\, 0} \sim 0.01$ and $a_\cc \lsim a_\eq$ and $a_\nr \gsim 10^{-2}$. This corresponds to the parameter range in \Eq{eq:parameter_bounds}.

\section{Density perturbations in the cannibal model}
\label{sec:cosmo}

With the thermal history and parameter space of the MC model determined we now study the effects of the cannibal fluid on density perturbations. In particular, we derive the suppression of the matter power spectrum (MPS) and solve for the region in parameter space with the correct amount of suppression to address the large-scale structure (LSS) discrepancy on $\sigma_8$. We start from the equations governing the evolution of the cosmological perturbations in the energy density and velocity of the different components of the Universe, focusing on the dark matter and cannibal fluids. In this Section we simply state the equations and study their solutions, first numerically and then analytically using simplifying approximations. We review the derivation of the perturbation equations in \App{appA}.

The equations for the cannibal and CDM perturbations in Fourier space are \cite{Ma:1995ey}:
\bea
	\dot\delta_\cc & = & -(1+w_\cc) \bl( \theta_\cc - 3 \dot\varphi \br) - 3 \mH \bl( c_s^2 - w_\cc \br)\delta_\cc \ , \label{eq:cannperts1a}\\
	\dot\theta_\cc & = & - \mH \bl( 1-3 c_s^2 \br)\theta_\cc + k^2 \bl( \psi + \frac{c_s^2}{1+w_\cc}\delta_\cc \br) \ ; \label{eq:cannperts1b} \\
	\dot\delta_\cdm & = & -\theta_\cdm + 3 \dot\varphi \ , \label{eq:cdmperts1a}\\
	\dot\theta_\cdm & = & - \mH \theta_\cdm + k^2 \psi \ , \label{eq:cdmperts1b}
\eea
where the dots represent derivatives with respect to conformal time $\eta$; $k$ is the Fourier momentum mode, $\mH \equiv aH = \dot a/a$, $\delta \equiv \delta \rho /\rho$ and $\theta$ are the density contrast and the velocity divergence perturbations, while $\varphi$ and $\psi$ are the scalar perturbations of the metric.\footnote{$\delta$ and $\theta$ are part of the stress-energy-momentum tensor $T_{\mu\nu}$ of their corresponding fluid, and their equations are obtained from the continuity equation $\nabla_\mu T^{\mu\nu} = 0$. For details see \App{appA}.} Finally, $w_\cc \equiv P_\cc/\rho_\cc$ is the equation of state of the $\phi$-sector, while $c_s^2 \equiv \dot P_\cc / \dot \rho_\cc = w_\cc - \frac{\dot w_\cc}{3\mH(1+w_\cc)}$ is the speed of sound of the $\phi$-fluid. Recall that during the cannibalistic phase $\rho_\cc \approx m n_\cc$ and $P_\cc \approx T n_\cc$ and therefore $w_\cc \approx T/m \sim 1/\log a$.

For the rest of this Section we make the following simplifications: {\it i.} ignore the baryons, adding their energy density to that of CDM, {\it ii.} ignore the anisotropic stress of the neutrinos, taking $\varphi = \psi$, and {\it iii.} add the neutrino energy density to that of the photons. Since we are only interested in the effects of cannibals on the MPS, we will compare the MPS in the theory with cannibals to the MPS in $\Lambda$CDM, evaluated today, and denote the ratio by $R(k)$:
\bea\label{eq:ratio}
	R(k) & \equiv & \frac{\mathrm{MPS}(k)_c}{\mathrm{MPS}(k)_\Lambda} \bigg\vert_\mathrm{today} = \frac{(\rho_\cdm \delta_\cdm + \rho_\cc \delta_\cc)_c^2}{(\rho_\cdm \delta_\cdm)_\Lambda^2} \bigg\vert_\mathrm{today} \nonumber\\
	& = & \bl( \frac{\delta_{\cdm,\,c}}{\delta_{\cdm,\,\Lambda}} + f_\cc \frac{\delta_\cc}{\delta_{\cdm,\,\Lambda}} \br)^2\bigg\vert_\mathrm{today} \ ,
\eea
where the index $c$ denotes the value in the theory with cannibals, while $\Lambda$ means $\Lambda$CDM. With the assumptions mentioned above, we solved \Eqst{eq:cannperts1a}{eq:cdmperts1b} numerically and calculated $R(k)$. We now describe the solutions for $\delta_\cc$ and $\delta_\cdm$, and the resulting $R(k)$.

\begin{figure}[!htbp]%
    \centering
    \includegraphics[width=0.55\textwidth]{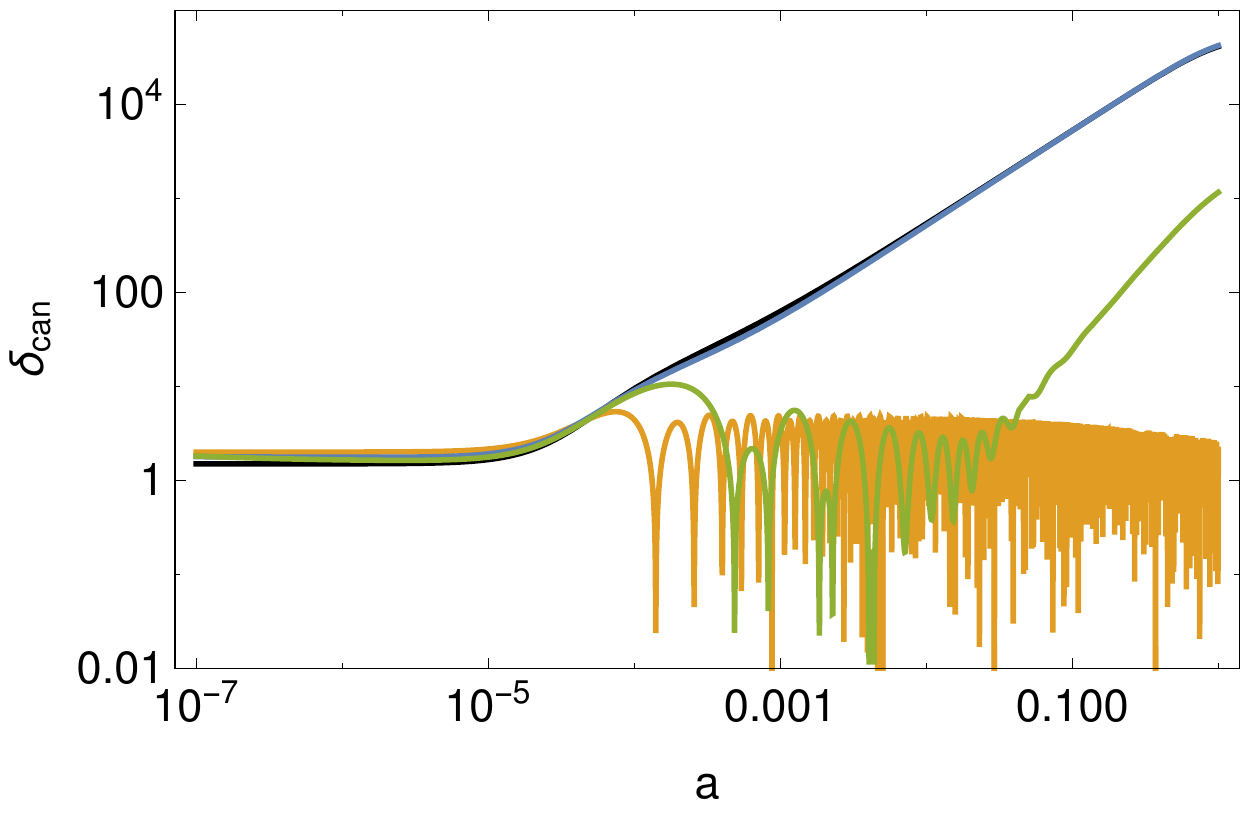}
    \caption{The cannibal perturbations for three choices of the MC model parameters, compared with the CDM perturbation from $\Lambda$CDM (black curve). The choice with early end of cannibalism is shown in blue, that with a late start of cannibalism in orange, while in green is that with the cannibalistic phase overlapping with matter domination. We have chosen $k=0.2h \, \mpc^{-1}$ with $h=0.68$; this corresponds to perturbations at the wave length which $\sigma_8$ is most sensitive to.}%
    \label{fig:perts}
\end{figure}
The evolution of the $\delta_\cc$ perturbations can be appreciated in \Fig{fig:perts}, for different choices of the parameters $m$ and $S_\cc$ of the MC model, having fixed $\alpha = 4 \pi$.
One choice of the parameters corresponds to early decoupling (blue curve), where the cannibalistic phase ends well before equality and the perturbations behave just like CDM. Another choice shows late cannibalization (orange line) in which the $\phi$-sector behaves just like radiation throughout most of the history of the Universe. In this case $\delta_\cc$ oscillates like radiation perturbations do. Since in this case the cannibalistic phase only starts when $\rho_\cc$ is already a negligible contribution to the total energy density, the cannibalism itself has no impact on the MPS. The green curve corresponds to the case of most interest: the cannibalistic phase overlaps with matter domination. The early part of the curve shows that cannibal perturbations perform acoustic oscillations after entering the horizon. The oscillations are due to the pressure term proportional to the speed of sound $c_s^2$ during cannibalism. Once the cannibalistic phase ends at $a_\nr$ the $\phi$ particles become non-relativistic and the speed of sound quickly drops $c_s^2 \approx T/m \sim a^{-2}$. This causes the $\delta_\cc$ perturbations to stop oscillating and to start growing by falling into the gravitational potentials sourced by the already clustered dark matter. This can be seen in the large-$a$ behavior of the green curve in \Fig{fig:perts}.

The cannibal fluid affects the perturbation equations for the CDM in two ways: through its contributions to the gravitational potential term $k^2 \psi$ in \Eq{eq:cdmperts1b} and through its contribution to the energy densities in the Hubble friction term $-\mH \theta_\cdm$ in \Eq{eq:cdmperts1b}; the $\dot\phi$ term in \Eq{eq:cdmperts1a} is negligible for the modes of interest.
Since $\delta_\cc$ oscillates and does not grow during the cannibalistic phase its contributions to the gravitational potential $\psi$ remain negligible and do not enhance the growth of CDM perturbations. On the other hand, the contribution of $\rho_\cc$ to the Hubble expansion rate during matter domination and therefore to the Hubble friction term is significant. The net effect,
no enhancement of the potential but more friction, is to slow the growth of CDM perturbations relative to $\Lambda$CDM. Thus the MPS is suppressed in theories with cannibals. This is the main result of our paper.

\begin{figure}[!htbp]%
    \centering
    \includegraphics[width=0.8\textwidth]{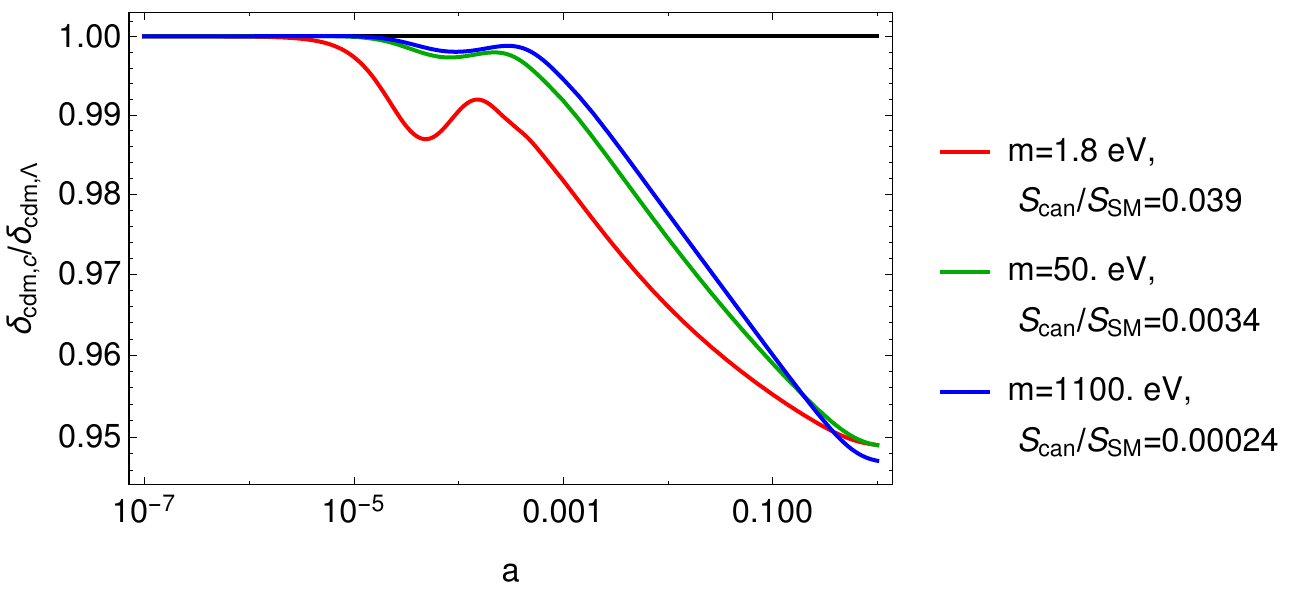}
    \caption{Evolution of the perturbation $\delta_\cdm$ for wave number $k=0.2h \, \mpc^{-1}$ in the presence of cannibals compared to its value in $\Lambda$CDM, for three different choices of model parameters. Models were chosen to give a $10\%$ suppression in the MPS today (\ie $R=0.9$). The three choices of $m$ and $S_\cc$ are also indicated as red, green, and blue points in \Fig{fig:mps_contour}.}%
    \label{fig:growthratio}
\end{figure}

\Fig{fig:growthratio} illustrates this result. We plot the ratio of $\delta_\cdm$ in the presence of cannibals to its value in $\Lambda$CDM as a function of the scale factor $a$ for the mode $k=0.2h \, \mpc^{-1}$. The three curves correspond to three models with parameters $m$ and $S_\cc$ chosen such that the MPS today for that mode is suppressed by 10\% (\ie $R(0.2h \, \mpc^{-1})=0.9$). Note that after some transitory behavior after the mode first enters the horizon the suppression increases monotonically during matter domination. This shows that the rate of growth in the presence of cannibals is smaller than in $\Lambda$CDM. This ratio behaves approximately like a power law in $a$, with a slight decrease of its slope which comes from the time dependence of $f_\cc \equiv \rho_\cc/\rho_\cdm$.

\begin{figure}[!htbp]%
    \centering
    \includegraphics[width=0.6\textwidth]{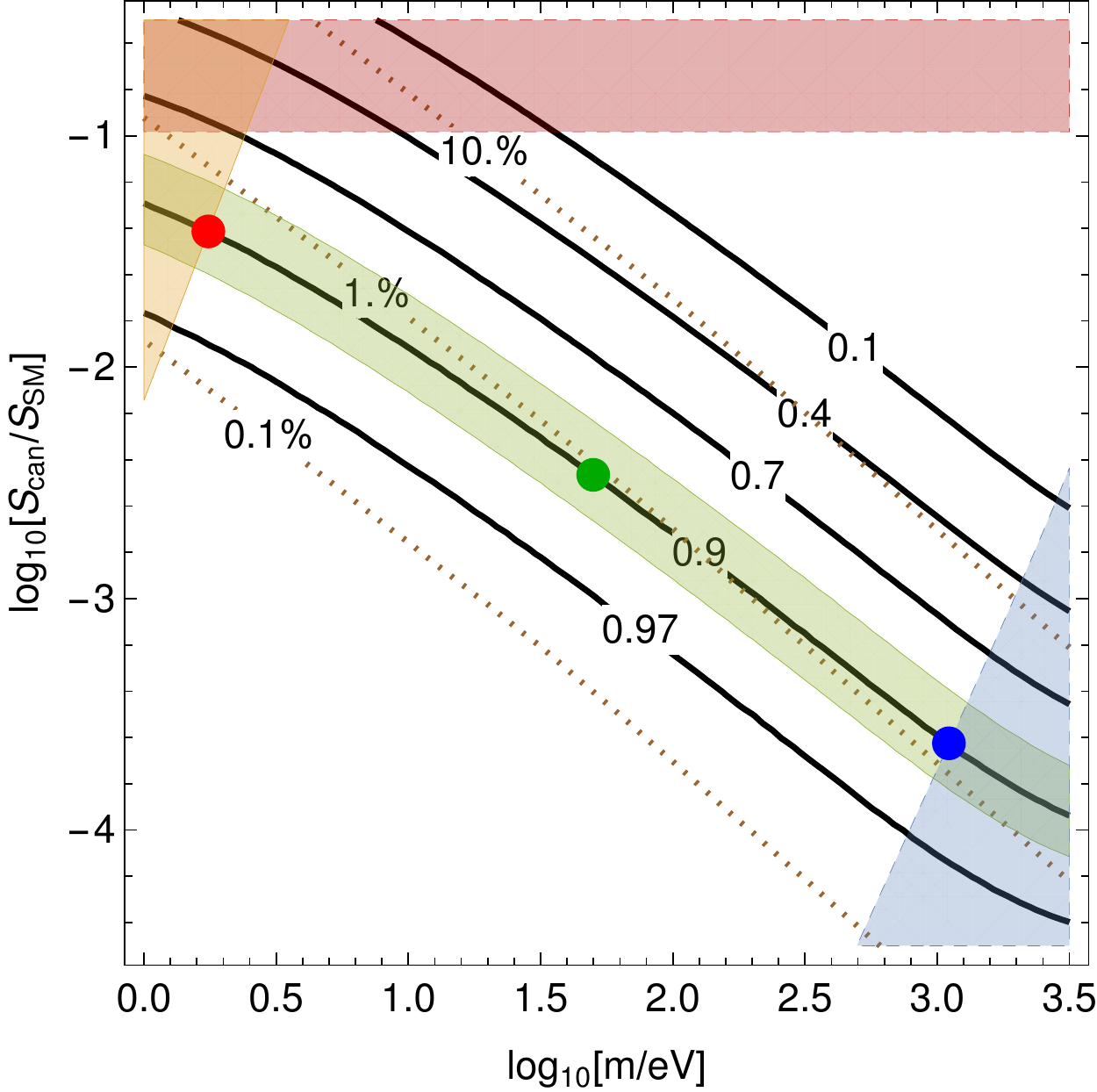}
    \caption{$m$ versus $S_\cc/S_\SM$ parameter space where $S_\SM = 2.2\times 10^{-11}\, \ev^3$ is the entropy in the Standard Model today. The black lines are contours of the ratio of the MPS in the presence of cannibal dark matter to that of $\Lambda$CDM. The brown dotted curves correspond to constant $f_{\cc,\,0}$. The green band is an estimate for the suppression that gives a $\sigma_8$ within 1$\sigma$ of the value quoted in \cite{Joudaki:2016mvz}. The orange region corresponds to MC models that enter the cannibalistic phase after matter-radiation equality, while the blue one corresponds to those for which cannibalism ends before $a=10^{-2}$. In red are those models whose $\rho_\cc$ contributes to $\Delta N_\eff \vert_\mathrm{BBN}>0.66$ \cite{Steigman:2012ve} when they are in their radiation phase. The red, green, and blue points correspond to the three choices of $m$ and $S_\cc$ in \Fig{fig:growthratio}.}%
    \label{fig:mps_contour}
\end{figure}

\Fig{fig:mps_contour} we show the $m$ - $S_\cc$ parameter space, with $S_\cc$ normalized to the entropy of the standard model today $S_\SM$. The black contour lines show $R(0.2h \, \mpc^{-1})$. In all the calculations for this plot we chose $\alpha = 4\pi$. We will study the (very small) dependence of the suppression $R(k)$ on the choice of $\alpha$ at the end of this Section. The brown dotted lines show the fraction $f_{\cc,\, 0}$ of cannibal dark matter today. The green band in \Fig{fig:mps_contour} represents the region of parameter space that yields a suppression in the value of the MPS today within 1$\sigma$ of the preferred value of $\sigma_8$ according to \cite{Joudaki:2016mvz}, about a 10\% suppression ($R(0.2h \, \mpc^{-1})=0.9$). We see that this roughly corresponds to $f_{\cc,\,0} \sim 1\%$. The orange region corresponds to the lower bound on $m$ we estimated in \Sec{sec:thermo}, made up of those parameter values for which $a_\cc > a_\eq$. Deep inside this region the $\phi$-fluid behaves just like radiation. The blue region corresponds to the upper bound also estimated in \Sec{sec:thermo}, for which $a_\nr < 10^{-2}$. Deep inside this region the $\phi$-fluid behaves like ordinary CDM. Finally, the red band corresponds to a region in parameter space in which the $\phi$-fluid would contribute too much radiation ($\Delta N_\eff > 0.66$) to the energy density of the Universe at the time of Big Bang Nucleosynthesis \cite{Steigman:2012ve}. However, as we will show in \Sec{sec:simplest} this constraint is relaxed in UV completions of the MC model because the energy density in radiation in the UV is reduced in such models.

The black contours showing the values for $R(k)$ were calculated for $\alpha=4\pi$. Since the value of $\alpha$ determines the scale factor at which the \threeto interactions decouple and cannibalism ends, we expect some dependence of the predicted MPS on $\alpha$. However, within the range of parameters in \Fig{fig:mps_contour} this dependence is very weak. The two main effects are that cannibal perturbations stop oscillating and start catching up to the dark matter perturbation after decoupling. If they have enough time to grow they can have a non-negligible impact on the MPS via the second term in \Eq{eq:ratio} and they contribute to the gravitational potential. However for the points that we are interested in the cannibal perturbations remain too small to be important. A numerically more significant effect is that when the cannibal fluid stops cannibalizing its energy density transitions from scaling like $1/(a^3 \log a)$ to $1/a^3$. Thus a model in which the $\phi$ particles stop cannibalizing earlier will have more energy density in cannibals and therefore more Hubble friction. This effect is somewhat more important but still small. For example, choosing $m$ and $S_\cc$ as for the blue dot in \Fig{fig:mps_contour} but choosing $\alpha=1$ and $\alpha=\infty$ (\ie no decoupling of the \threeto interactions) we obtain $R=0.92$ and $R=0.902$ for the MPS ratio respectively, a very small effect.

Having shown that the presence of cannibals suppress the MPS by numerically solving the equations for the perturbations, we devote the rest of this Section to understanding this result from \Eqst{eq:cannperts1a}{eq:cdmperts1b}. We will only be interested in $k$ modes which are well inside the horizon during matter domination, \ie modes for which $k \gg 1/ \eta_\eq \sim 0.01 \mpc^{-1}$.

Let us start with the cannibal perturbations. For modes deep inside the horizon the gravitational potential is approximately constant so that we can ignore derivatives of $\psi$. In addition, we can use $w_\cc \ll 1$, $c_s^2 \ll 1$ to drop all subleading terms in \Eqs{eq:cannperts1a}{eq:cannperts1b}. Then taking the second derivative of $\delta_\cc$ and substituting \Eq{eq:cannperts1b} into \Eq{eq:cannperts1a} yields:
\beq\label{eq:cannperts2}
	\ddot \delta_\cc + \mH \dot \delta_\cc + k^2c_s^2 \delta_\cc = - k^2 \psi  \ ,
\eeq
where the term on the right-hand-side is the solution of the Poisson equation
\beq\label{eq:poisson}
-k^2 \psi = \frac{3}{2}\, \frac{a^2}{3 M_\Pl^2} \sum_i \rho_i \delta_i \ .
\eeq

Anticipating that the CDM contribution dominates the sum during matter domination (duh!), and that perturbations in the CDM fluid grow linearly, $\delta_\cdm \sim a$, one sees explicitly that $\psi$ is constant during
matter domination. Thus \Eq{eq:cannperts2} is a simple harmonic oscillator with friction and the gravitational potential corresponds to a constant shift of the zero point. The solutions are oscillatory as long as $k c_s > \mH \sim 1/\eta$, \ie as long as the $k$-modes are small compared to the sound horizon, $2\pi/k \ll c_s \eta$. Recalling that $c_s^2 \approx w_\cc \approx T/m \sim 1/\log a$ for cannibals and $\eta \sim \sqrt{a}$ during matter domination it is clear that modes which are inside the Hubble horizon also enter the growing sound horizon $c_s \eta \sim \sqrt{a/\log a}$ and oscillate. However, once cannibalism ends, $c_s \sim 1/a$.  Then the sound horizon $c_s \eta \sim 1/\sqrt{a}$ shrinks and the mode eventually exits the sound horizon, stops oscillating and starts growing. However, for the region of parameter space that we are interested in the cannibal perturbations do not catch up to the CDM perturbations, thus justifying our approximation to only keep the CDM term in the gravitational potential, \Eq{eq:poisson}.

We now turn our attention to the CDM perturbations. Following the same procedure as before, combining \Eqs{eq:cdmperts1a}{eq:cdmperts1b} gives:
\beq\label{eq:cdmperts2}
	\ddot \delta_\cdm + \mH \dot \delta_\cdm + k^2 \psi =0 \ ,
\eeq
where $\psi$ is given by \Eq{eq:poisson} but only keeping the CDM contribution $\rho_\cdm \delta_\cdm$ in the sum. Using this, rewriting the Hubble parameter in terms of the energy density during matter domination $\rho_\tot \simeq \rho_\cdm + \rho_\cc$, and changing variables from $\eta$ to $a$ we can write:
\beq\label{eq:cdmperts3}
	(\rho_\cdm + \rho_\cc) \, a^2 \delta_\cdm'' + \frac{3}{2}\bl( \rho_\cdm + \rho_\cc \br) a \delta_\cdm' - \frac{3}{2}\rho_\cdm \delta_\cdm = 0 \ .
\eeq

Were it not for the cannibals, this would be the M\'{e}sz\'{a}ros equation during matter domination, whose growing solution is $\delta_\cdm \sim a$. \Eq{eq:cdmperts3} shows that cannibal dark matter increases the Hubble friction ($\delta_\cdm'$ term) felt by the CDM perturbations but does not contribute to the gravitational pull from the Poisson term. This explains the smaller rate of growth of $\delta_\cdm$ we discovered in our numerical solutions.

To get a rough idea of what this change in the growth rate is let us further simplify \Eq{eq:cdmperts3} by taking $\rho_\cc/\rho_\cdm \ll 1$ and dividing by $\rho_\cdm + \rho_\cc$ to arrive at \Eq{eq:cdmperts}. This is easily integrated in an approximation where we
neglect the slow $\log a$ dependence of $\rho_\cc$. In fact, this equation for the growth of perturbations without the $\log a$ dependence applies to a model with CDM and a subdominant component of dark plasma \cite{Chacko:2016kgg,Buen-Abad:2017gxg}. The solution for the growing mode  is the power law $\delta_\cdm \sim a^{1- \frac{3}{5} \rho_\cc/\rho_\cdm}$ \cite{Lesgourgues:1519137,Chacko:2016kgg,Buen-Abad:2017gxg}, a growth rate smaller than the linear one from the usual M\'{e}sz\'{a}ros equation. For the decaying mode, one finds $a^{-\frac32+\frac{3}{5} \rho_\cc/\rho_\cdm}$. In the cannibal case the exponent is a slowly varying integral function of $f_\cc$ that depends on $a$ (because of the slow logarithmic decay of $f_\cc$), which explains the change in the slope of the suppression we saw in \Fig{fig:growthratio}.

\section{Natural UV completions from secluded gauge sectors}
\label{sec:simplest}

In this Section we discuss our favorite UV completion of the MC model, a simple non-Abelian ``pure-glue" gauge sector which confines at low energies and produces canniballistic glueballs. 

Consider an $SU(N)$ gauge theory with no light matter fields. Such a theory has two marginal operators, the gauge kinetic term
\bea
-\frac{1}{4g_D^2} F_{\mu\nu}^2 
\label{eq:kineticterm}
\eea
and the CP-violating $\theta F\tilde F$ term. We set $\theta=0$ mostly because it makes no qualitative difference but also because it is zero if the dark sector preserves CP.  All other operators as well as couplings to the SM are irrelevant (in the sense of their scaling with energy) and therefore do not impact the confining dynamics and cannibalism. The dark sector could be coupled to the SM in the UV by heavy matter fields which are charged under both the SM and dark $SU(N)$ gauge group. Then it would be natural for the two sectors to have a common temperature in the UV. However if inflation and reheating occur at temperatures below the coupling of the two sectors or if there is a phase transition or there are heavy particles with associated entropy production then the two sectors may end up with very different temperatures. We take the temperature of the cannibal sector to be a free parameter $T$. 

Assuming that the $SU(N)$ gauge coupling in the UV is not too small the coupling runs strong in the IR and the theory confines at temperatures below some scale $\Lambda_c$. The confining gauge theory has a spectrum of stable glueball states with varying spin and parity quantum numbers \cite{Cornwall:1982zn,Morningstar:1999rf,Kribs:2009fy,Forestell:2016qhc}. The most important of these glueballs for cosmology is the lightest glueball $\phi$ with mass $m \sim \Lambda_c$ which is a parity even scalar and carries no conserved quantum number. It has number-changing interactions and its low energy effective description is the Lagrangian \Eq{eq:lagrange} plus higher-dimensional operators of the form $\phi^n/m^{n-4}$. The important parameters of this low-energy theory are the glueball mass $m$ and the entropy in the glueballs $S_\cc$. There is also a dependence on the coupling $\lambda$ which determines the end of the cannibalism phase when $\Gamma_{32} = H$. For a strongly coupled $SU(N)$ theory naive dimensional analysis predicts $\lambda \simeq 4 \pi/\sqrt{N}$. Changing this coupling by a factor of 2 would change the duration of the cannibalism phase by 1/2, see \Eq{eq:duration2}, this has very little impact on the cosmology.
Note that the number density of heavier glueballs $\phi_H$ is exponentially suppressed relative to $\phi$ at low temperatures even if they are stable because they can efficiently annihilate $\phi_H + \overline \phi_H \rightarrow \phi + \phi$. 

Since the $\phi$ particles have no conserved quantum number they are unstable to decay. $\phi$ has no other particles to decay to in the dark sector but it can decay to gravitions or SM particles through higher dimensional operators. For example, the width to decay into gravitons is roughly $m^5/M_{Planck}^4 \sim 10^{-108}\, \ev [m/\ev]^5$. This is much smaller than the Hubble constant today for the masses we consider. In fact, even decays mediated by a dimension 6 operator suppressed by a scale of 1 GeV are too slow to be cosmologically relevant for $m \sim 1$ eV. This justifies treating the $\phi$ particles as stable.

This completes our description of the UV completion of the MC model. In most of the interesting parameter space, \Fig{fig:mps_contour}, the UV completion is not needed for the computation of the MPS. This is because either {\it i.} the confining transition happens well before matter-radiation equality and the energy density in the cannibal sector is negligible during and before the transition or {\it ii.} because the confining transition happens well after matter-radiation equality. In the latter case the cannibal Sector is ``gluon" radiation well into matter domination and its energy density redshifts to being negligible before cannibalism even starts. Thus only in models where the confinement transition happens close to matter-radiation equality (red dot in \Fig{fig:mps_contour}) is the UV completion needed for the computation of the MPS. We study this special case in the remainder of this Section.

Computing the cosmological evolution of the cannibal fluid through the confining phase transition exactly is very difficult as one would have to solve for the dynamics of a strongly coupled thermal gauge theory \cite{Witten:1984rs}. We take a simplified approach and match the UV theory with $N^2-1$ weakly interacting gluons onto the confined theory of the lightest glueball $\phi$. This matching depends on the size of the gauge group, $N$, the details of the phase transition (it can be 1st or 2nd order), the glueball spectrum, and couplings in the strongly coupled regime. It is believed that for $N=2$ the phase transition is 2nd order so that entropy is conserved in the phase transition, for $N=3$ it is probably weakly 1st order and for higher $N$ strongly 1st order \cite{Lucini:2003zr,Lucini:2005vg}. Note that in the presence of extra matter with mass near the confinement scale the order of the phase transition can change. Thus we treat the order of the phase transition as an additional uncertainty. In the case of a strongly 1st order phase transition the gluon plasma super-cools below the confinement scale before critical bubbles of the confined phase appear. In such a scenario, the entropy increases during the phase transition, and because of the super-cooling only the lightest glueballs are abundant after the phase transition.

The model dependence due to unknown physics of the phase transition enters into the matching onto the UV theory. A single IR theory which is specified by giving $S_c$ and $m$ can match onto different UV theories, with different values of $N$ and possibly different  phase transitions. To study the sensitivity of the MPS predictions to this model-dependence we look at two simplified cases: a smooth 2nd order phase transition with conserved entropy and a simplified glueball spectrum and a very strongly 1st order phase transition with a jump in entropy and temperature (see for example \cite{Megevand:2016lpr}).

To model the 2nd order phase transition we assume that the full theory is described by $g_*=2(N^2-1)$ bosonic degrees of freedom. The lightest, $\phi$, has mass $m$, and the others have a common mass $M$ which we vary from 1.25$m$ to 3$m$. We also assume entropy conservation and that the theory remains in chemical and thermal equilibrium throughout the phase transition. Then all distributions are simply given by Boltzmann distribution functions for the $2(N^2-1)$ degrees of freedom. In the UV, when all masses can be ignored, this reproduces the physics of free $SU(N)$ gluons. In the transition region where $T\sim m$ the ``heavy glueballs" of mass $M$ pair annihilate into the lightest glueballs $\phi$. And in the IR, when the temperature drops below $m$, only the cannibals remain.
\begin{figure}[!htbp]%
    \centering
    \includegraphics[width=0.4\textwidth]{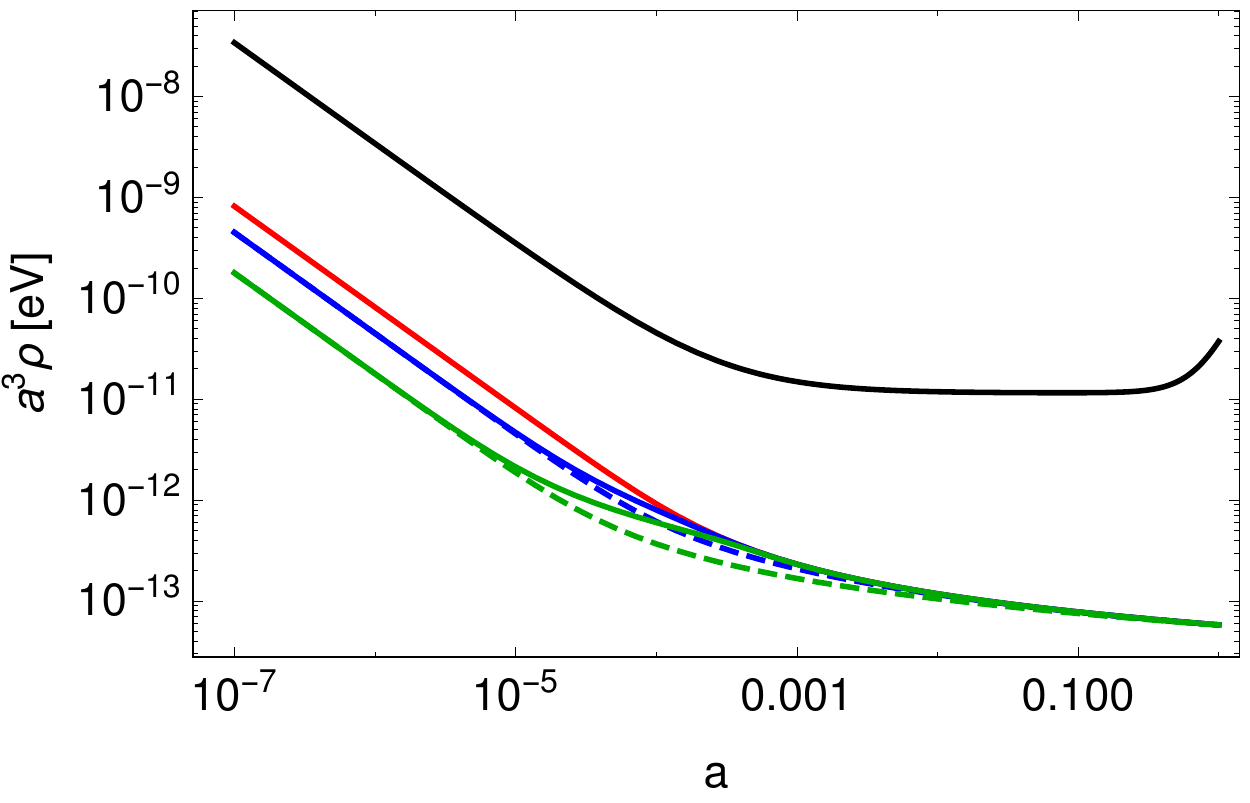}
    \includegraphics[width=0.55\textwidth]{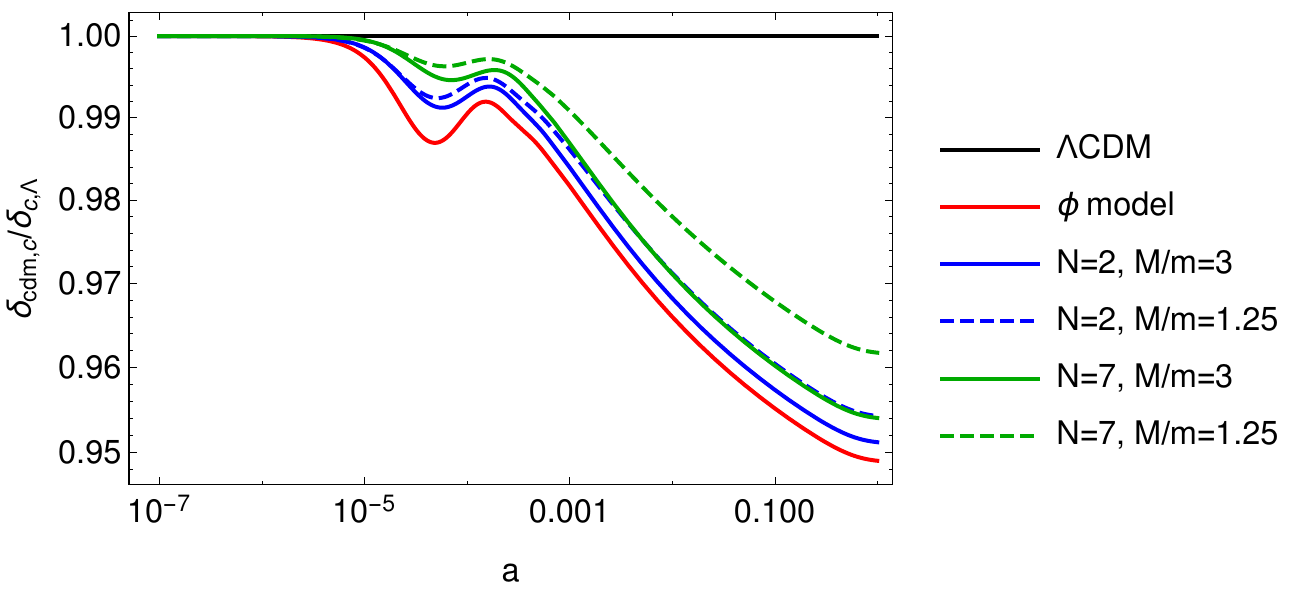}
    \caption{Plots of $a^3\rho_\cc$ (left) and $\delta_\cdm$ ratio (right) for different UV completions with 2nd order phase transitions compared to the MC model that gives $R(0.2h \, \mpc^{-1})=0.9$ and $a_\cc = a_\eq$ (\ie $m=1.8 \, \ev$, $S_\cc/S_\SM = 0.04$, corresponding to the red dot in \Fig{fig:mps_contour}). The black lines correspond to $\Lambda$CDM while the colored lines to the cannibal fluid in different models. The energy densities are continuous in $a$, because entropy is conserved throughout the transition. For the different UV completions we vary the number of UV degrees of freedom $g_*=2(N^2-1)$ corresponding to dark gauge groups $SU(N)$ as well as the masses $M$ of the heavier glueball states. The MPS ratio is less suppressed, from $R=0.905$ to $R=0.925$ for the $N=2$ and $M/m=3$ (solid blue) and $N=7$ and $M/m = 1.25$ (dashed green) lines respectively.}%
    \label{fig:uv2}
\end{figure}

For the very strongly 1st order phase transition we match a UV theory of $N^2-1$ massless gluons onto the IR theory with a jump in entropy at a scale factor $a_\cc$. We choose the matching scale factor such that the temperature evaluated in the IR theory (the theory of the cannibal $\phi$) equals $m/3$ at the matching scale. There we match onto the UV theory with $g_*=2(N^2-1)$ massless bosonic degrees of freedom and a jump in entropy (increasing from the UV to the IR) by a multiplicative factor which we vary from 1 to 2. The discontinuity in degrees of freedom and entropy at the matching point also implies a discontinuity in other background quantities. 
\begin{figure}[!htbp]%
    \centering
    \includegraphics[width=0.4\textwidth]{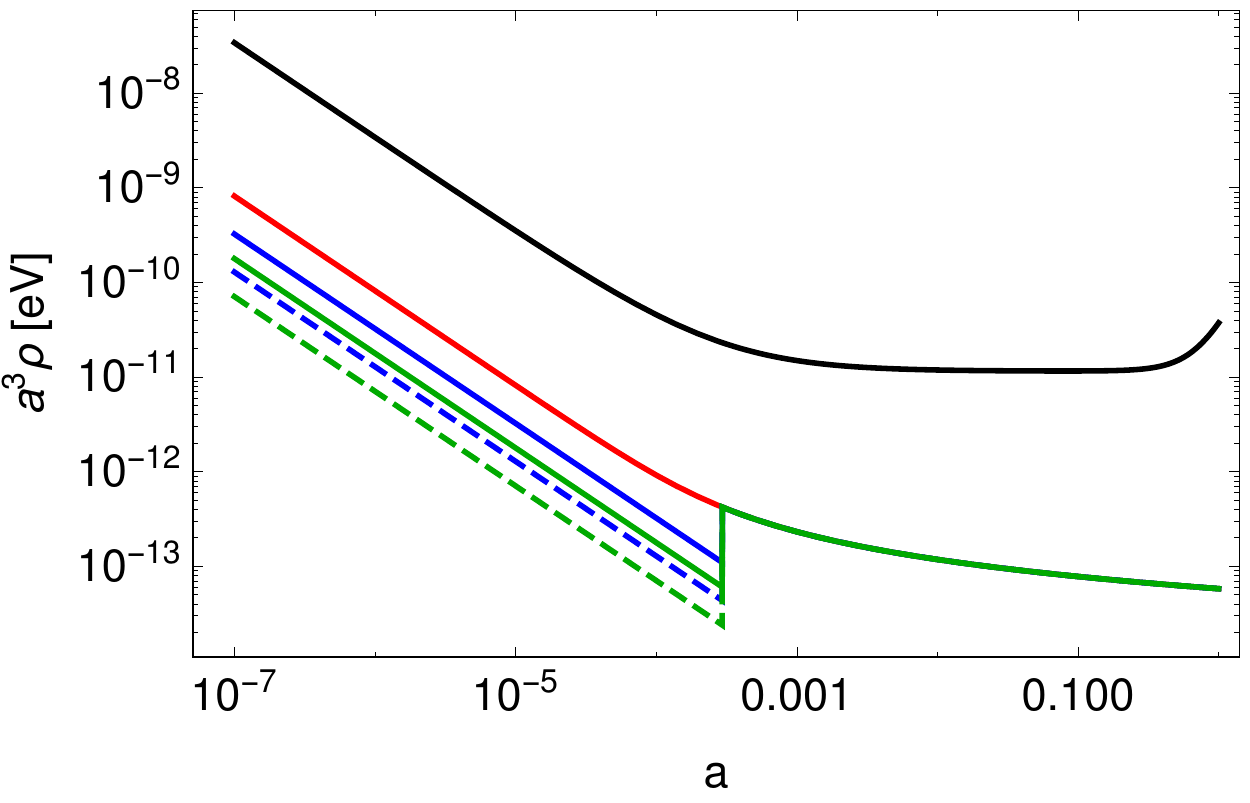}
    \includegraphics[width=0.53\textwidth]{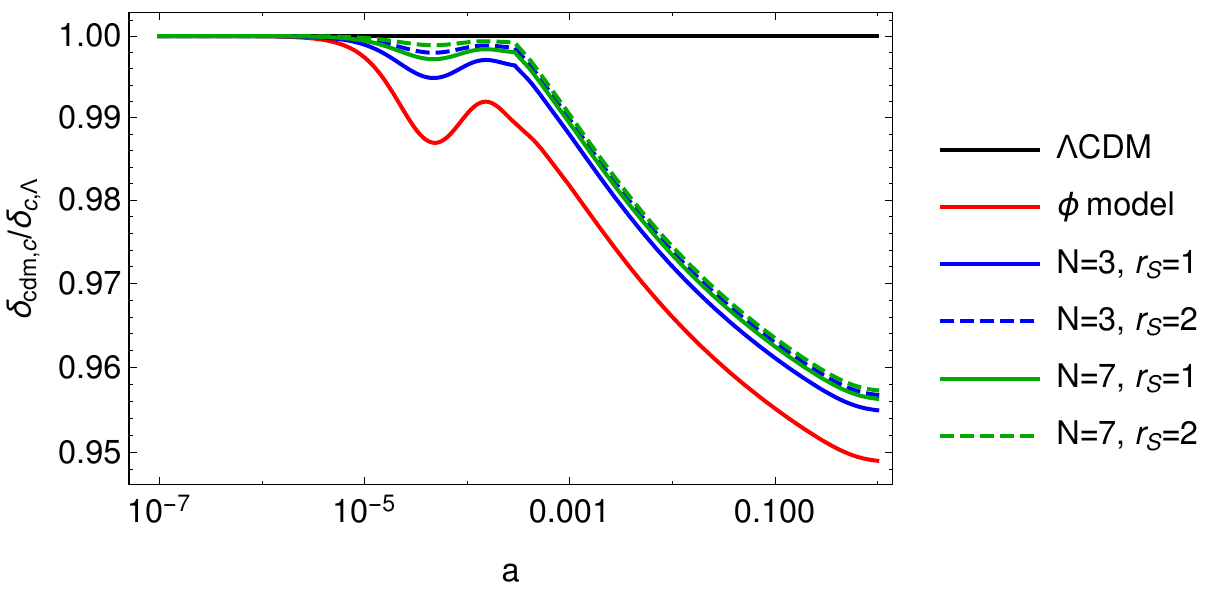}
    \caption{Plots of $a^3\rho_\cc$ (left) and $\delta_\cdm$ ratio (right) for different UV completions with 1st order phase transitions compared to an MC model that gives $R(0.2h \, \mpc^{-1})=0.9$ and $a_\cc = a_\eq$ (\ie $m=1.8 \, \ev$, $S_\cc/S_\SM = 0.04$, corresponding to the red dot in \Fig{fig:mps_contour}). For the different UV completions we vary the strength of the discontinuity in the entropy at the matching scale $a_\cc$, parametrized by the multiplicative factor $r_S\equiv S_\cc/S_{UV}|_{a_\cc}$ and the size of the UV gauge group $SU(N)$. The MPS ratio is less suppressed, from $R=0.912$ to $R=0.916$ for $N=3$ and $r_S = 1$ (solid blue) and $N=7$, $r_S=2$ (dashed green) lines respectively.}%
    \label{fig:uv1}
\end{figure}

\section{Conclusions}
\label{sec:conc}

We have studied the possibility that a subdominant component of the dark matter might posses a cannibalistic phase. If this phase overlaps with matter domination then the most significant impact is on the matter power spectrum.  This is particularly interesting because there is 2-3 $\sigma$ tension in direct observations of the matter power spectrum a 8 Mpc$^{-1}$ scales with the matter power spectrum inferred from $\Lambda$CDM and the precision fit to the CMB data from Planck \cite{Heymans:2013fya,Joudaki:2016mvz,Ade:2015fva,Ade:2013lmv,Kohlinger:2017sxk,Joudaki:2017zdt}. Even if one dismisses the hints for new physics from this source observations of the matter power spectrum are going to improve significantly in the coming years with much more precision on the full spectral shape (as a function of $k$) expected. Thus we find it interesting to explore what impact different types of new physics may have on the shape of the matter power spectrum. 

The simple cannibal model of \Eq{eq:lagrange} has three parameters which characterize its fluid description. Given our preference for strongly coupled UV completions of the simple model, one of them is more or less fixed: $\alpha \sim 4\pi$. Its significance is to determine the scale factor at which the \threeto interactions decouple and the $\phi$ particles stop cannibalizing and turn into cold dark matter. Smaller values of $\alpha$ would lead to a shorter period of cannibalization. The other two parameters characterizing the cannibal fluid are its entropy $S_\cc$ and the mass $m$ of the cannibal particle. We conclude this Section with two plots which show the impact of these two parameters on the predicted matter power spectrum shape. 

\begin{figure}[!htbp]%
    \centering
    \includegraphics[width=0.7\textwidth]{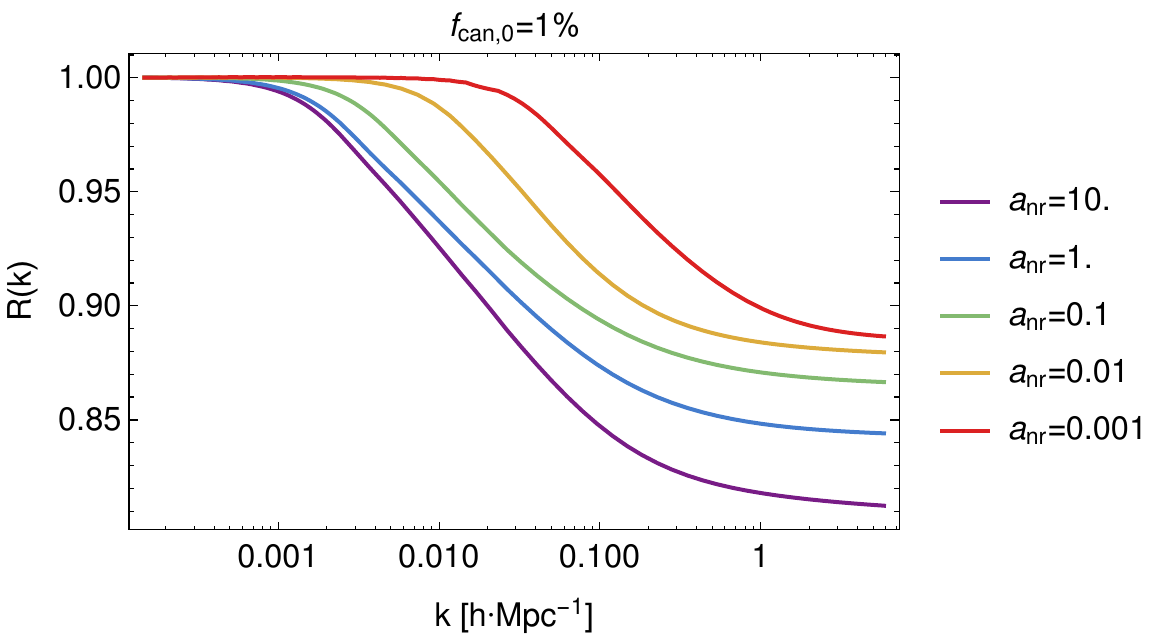}
    \caption{MPS ratio $R(k)$ for different values of $a_\nr$ and fixed $f_{\cc,\,0} =1\%$, normalized such that there is a $1\%$ of extra CDM in the $\Lambda$CDM theory in order to cancel some background effects. The later $a_\nr$ is, the more small $k$ modes can enter the sound horizon and present cannibal acoustic oscillations, suppressing the MPS. Note that even though $f_{\cc,\,0}$ is fixed the large-$k$ MPS suppression is not the same for different $a_\nr$. This is because if cannibalism is still happening during matter domination (\ie $a_\nr > a_\eq$) then $f_\cc$ is bigger earlier in the Universe, because of its logarithmic scaling, and this enhances the suppression.}
    \label{fig:mps_anr}
\end{figure}

\Fig{fig:mps_anr} shows the dependence of the MPS on the decoupling scale $a_\nr$. For fixed $\alpha = 4\pi$ we have roughly $a_\nr \sim 10^5\, a_\cc \sim 10^6\, S_\cc^{1/3}/m$, thus $a_\nr$ depends on the ratio of $S_\cc^{1/3}$ and $m$. This scale is when cannibalism stops, therefore any wave mode $k$ which enters the (sound) horizon after this scale cannot be affected by the cannibal fluid oscillations and will take on the same value as in $\Lambda$CDM. Thus $a_\nr$ can be understood to determine the smallest values of $k$ which are suppressed by cannibalism. Therefore changing the ratio $ S_\cc^{1/3}/m$ which changes $a_\nr$ is equivalent to shifting the MPS suppression curve in the horizontal $k$ direction. For the purposes of this plot we fixed the fraction of the energy density in the cannibal fluid today relative to the ordinary dark matter energy density to $f_{\cc,\,0} =0.01$ for all models. The $\Lambda$CDM reference power spectrum which we compare to (the denominator of $R$) has $1\%$ of additional dark matter instead of the cannibal fluid so that all models being compared have the same value of $H_0$. This removes the background effect of the additional energy density in the cannibal fluid.

\begin{figure}[!htbp]%
    \centering
    \includegraphics[width=0.8\textwidth]{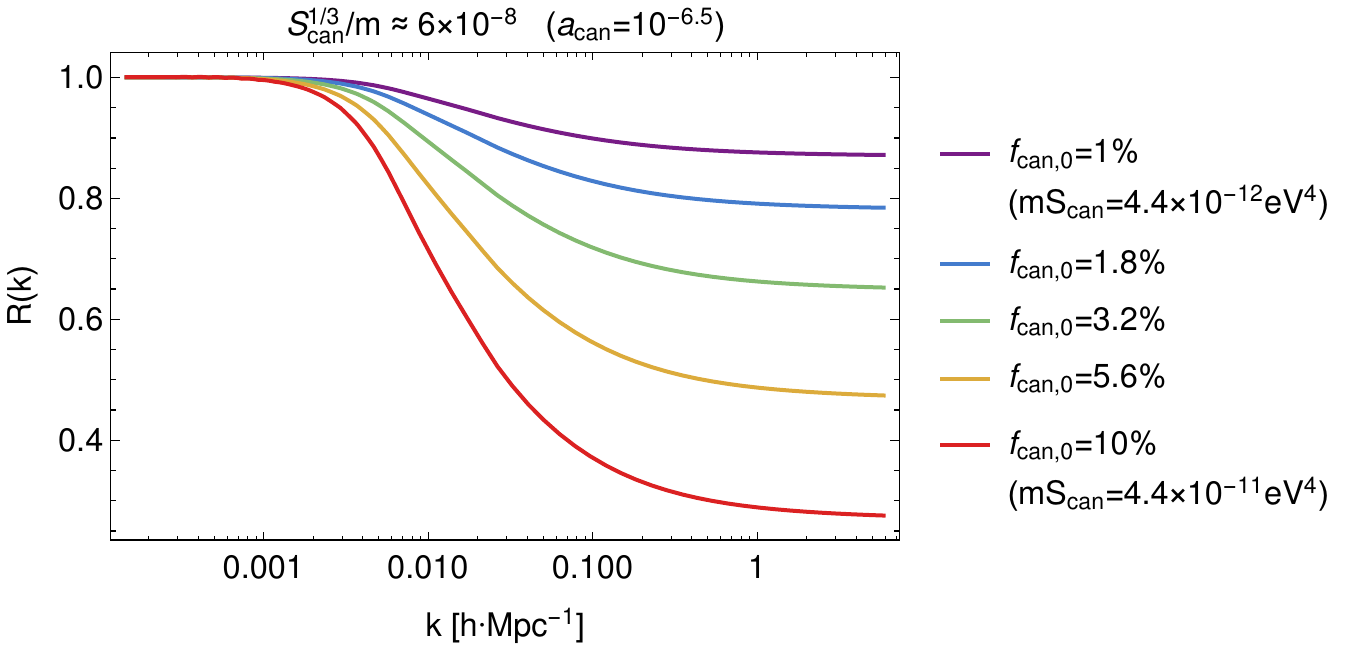}
    \caption{MPS ratio $R(k)$ for different values of the product $m S_\cc$ but fixed ratio $S_\cc^{1/3}/m$ (\ie fixed $a_\cc$). This corresponds to different fractions $f_{\cc,\,0}$ of cannibals, from $1\%$ (purple) to $10\%$ (red). We have normalized $R$ such that there is a corresponding extra amount of CDM in the $\Lambda$CDM theory, in order to cancel out some background effects. With a fixed $a_\nr$ it is clear that the same $k$ modes are suppressed, but the amount of suppression is dialed by $f_{\cc,\,0}$.}%
    \label{fig:mps_mS}
\end{figure}

\Fig{fig:mps_mS} shows the dependence on the orthogonal combination of parameters. \ie varying $S_\cc^{1/3}$ and $m$ while holding their ratio fixed. This keeps the scales in $k$ at which the suppression occurs fixed but it changes the overall energy density in the cannibal fluid and therefore changes mostly the amplitude of the suppression. 

Note that this second dependence is similar to that of the MPS on neutrino mass \cite{Lesgourgues:2006nd}. However the smallest $k$ affected by non-zero neutrino masses is constrained to within a factor of a few of $k_{NR} \sim  0.01\, \mpc^{-1}$ whereas for cannibals the onset of the suppression in the MPS can lie anywhere within $k\sim 0.001 - 0.1\, \mpc^{-1}$ (see \Fig{fig:mps_anr}). 

Finally, we wish to mention the other ``anomaly" in cosmological precision fits: the discrepancy between the value of $H_0$ inferred from the Planck CMB data (and BAO) within $\Lambda$CDM and the direct measurement of $H_0$ from \cite{Riess:2016jrr,Bonvin:2016crt}. To see if cannibals could also help with this anomaly while remaining consistent with everything else would require a global fit of the cannibal model.  

\section*{Acknowledgments}
\label{sec:ack}

We wish to thank Ami Katz and William Sheperd for discussions on phase transitions and David E. Kaplan and Neil Weiner for suggestions which lead to the birth of this project. We are also thankful to Julien Lesgourgues and Deanna Hooper for catching a typo in one of our equations. The work of MB and MS is supported by DOE grant DE-SC0015845. The work of RE was supported by Hong Kong University through the CRF Grants of the Government of the Hong Kong SAR under HKUST4/CRF/13. RE acknowledges the members of the particle theory group at Boston University for their very warm hospitality and support during her visit to BU when most of this work was done. MS appreciated the hospitality of the IAS at HKUST where his stay was just amazing.


\appendix

\section{Cannibal equations from the Boltzmann equation}
\label{appA}

In this Appendix we derive the equations that describe the $\phi$-fluid from the statistical description of its particles, and show that the self-interactions of the fluid do not appear. In particular we obtain \Eqs{eq:cannperts1a}{eq:cannperts1b}.

\subsubsection*{The Boltzmann equation}

The fluid description for a fluid $\phi$ can be derived from its one-particle distribution function $f(\mathbf{p}, \mathbf{x}, \eta)$ for a $\phi$ particle of mass $m$, momentum $\mathbf{p}$, energy $E = \sqrt{\mathbf{p}^2 + m^2}$, and position $\mathbf{x}$, at a conformal time $\eta$. To zeroth order the Universe is isotropic and homogeneous, and thus it is described by the Friedmann-Lema\^{i}tre-Robertson-Walker metric. Therefore, its distribution function is similarly homogeneous and isotropic to zeroth order. Expanding around this background to first order in small inhomogeneities:
\beq\label{eq:A1}
  f(\mathbf{p}, \mathbf{x}, \eta) = \f0(p,\eta) \bl( 1 + \Psi ( \mathbf{p},\mathbf{x}, \eta) \br) \ ,
\eeq
where $\Psi$ is the small perturbation in the distribution function that codifies the inhomogeneities and anisotropies of the $\phi$-fluid. Since we are ultimately interested in the matter power spectrum (MPS), we Fourier transform $f$ and $\Psi$ with regards to $x$ to arrive at equivalent expressions for the Fourier transforms with $\mathbf{x} \rightarrow \mathbf{k}$.

From the distribution function one can calculate the quantities that describe the $\phi$-fluid (\cite{Ma:1995ey}):
\bea
\overline{n}(\eta) + \delta n (\mathbf{k}, \eta) & \equiv & \int \text{D} p \,\, \f0 ( 1 + \Psi ) \ , \label{eq:A2} \\
\overline{\rho}(\eta) (1 + \delta (\mathbf{k}, \eta)) & \equiv & \int \text{D} p \,\, \f0 ( 1 + \Psi ) \ E \ , \label{eq:A3} \\
\overline{P}(\eta) + \delta P(\mathbf{k}, \eta) & \equiv & \int \text{D} p \,\, \f0 (1 + \Psi) \ \frac{p^2}{3E} \ , \label{eq:A4} \\
(\overline{\rho}+\overline{P}) \theta(\mathbf{k}, \eta) & \equiv & \int \text{D} p \,\, \f0 \Psi \ (i \, \mathbf{k} \cdot \mathbf{p}) \ , \label{eq:A5}
\eea
with $\text{D} p \equiv g \frac{\text d^3 \mathbf{p}}{(2 \pi)^3}$ being the phase space element, assuming each $\phi$ particle has $g$ degrees of freedom. From now on we take $g=1$, as in our MC model. For brevity we also drop the overline denoting a background (average) quantity.

The Boltzmann equation describes the evolution of the distribution function:
\beq\label{eq:A6}
  \dot{f} + i k \frac{p}{E} \hat{p} \cdot \hat{k} \ f +  p \frac{\partial f}{\partial p} \bl( -\mathcal{H} + \dot{\varphi} - i k \frac{E}{p} \hat{p} \cdot \hat{k} \ \psi \br) = \frac{a}{2E} (1+\psi) C[f] \ ,
\eeq
where $\psi$, $\varphi$ are the Newtonian gauge scalar perturbations of the metric, $a$ is the scale factor, and the dots are derivatives with respect to conformal time. The left hand side describes the free evolution of the distribution function, whereas the right hand side encodes the change in $f$ due to collisions,  and it is appropriately called the collision operator $C[f]$.

Since we are interested in identifying our $\phi$-sector with our cannibal fluid, we will assume that the particles have \twoto and \threeto interactions, and therefore write $C[f] \equiv C_{22}[f] + C_{32}[f]$. The change in $f$ arising from collisions must be proportional to the distributions of the particles involved, as well as to the amplitude squared of the interactions. We can then write the collision term for $f_1 \equiv f(\mathbf{p}_1, \mathbf{k}, \eta)$ as:
\bea
  C_{22}[f_1]  =&  &\!\!\!\!\!\! \int \ \mathrm{d}\Pi_2 \mathrm{d}\Pi_3 \mathrm{d}\Pi_4 \,\, F_{22} \ , \label{eq:A7} \\
  C_{32}[f_1] =&  &\!\!\!\!\!\! \int \ \mathrm{d}\Pi_2 \mathrm{d}\Pi_3 \mathrm{d}\Pi_4 \mathrm{d}\Pi_5 \,\, F_{32} \ , \label{eq:A8} \\
\quad F_{22}  \equiv& &\!\!\!\!\!  \frac{1}{1!2!}\bl(-f_1 f_2 + f_4 f_3 \br)\vert \mathcal{M}_{22}(p_1,p_2;p_3,p_4) \vert^2 (2 \pi)^4 \delta^4 \bl( p_1 + p_2 - p_3 - p_4 \br)  \ , \label{eq:A9} \\
  F_{32}  \equiv &&\!\!\!\!\!  \frac{1}{2!2!}\bl(-f_1 f_2 f_3 + f_4 f_5 \br) \vert \mathcal{M}_{32}(p_1,p_2,p_3;p_4,p_5) \vert^2 (2 \pi)^4\delta^4\! \bl( p_1\!+\!p_2 \!+\! p_3\!-\! p_4\!-\! p_5 \br) \nonumber\\
  +&& \!\!\!\!\!\!  \frac{1}{1!3!}\bl(f_4 f_2 f_3 - f_1 f_5 \br) \vert \mathcal{M}_{32}(p_4,p_2,p_3;p_1,p_5) \vert^2 (2 \pi)^4 \delta^4\! \bl( p_4\!+\!p_2 \!+\! p_3\!-\! p_1\!-\! p_5 \br)\! , \label{eq:A10}
\eea
where $\mathrm{d} \Pi_i \equiv \frac{\text{D} p_i}{2 E_i}$ is the Lorentz-invariant momentum space element, while, for example, $\vert \mathcal{M}_{32}(p_1,p_2,p_3;p_4,p_5) \vert^2 $ is the $3 \rightarrow 2$ scattering amplitude squared, and we have used that the matrix element squared is invariant under permutation of the identical initial particles, final particles, and time reversal (swap of initial and final states). Note that for brevity we have not included the Bose enhancement factors $(1+f_i)$, but this does not affect any of our arguments. When the $\phi$ particles are non-relativistic the Bose enhancement reduces to 1.

For the \twoto scattering factor in \Eq{eq:A9} we have labeled the momenta as $\mathbf{p}_1 + \mathbf{p}_2 \leftrightarrow \mathbf{p}_3 + \mathbf{p}_4$. A plus sign denotes a ``gain'' in the distribution function of $f_1$ because of the appearance of a particle with momentum $\mathbf{p}_1$, while a minus sign denotes a ``loss'' of the same, due to the inverse process. The symmetry factors correspond to the different permutations of the initial and final state particles, once $\mathbf{p}_1$ has been selected.

For the \threeto scattering in \Eq{eq:A10} there are two places for $\mathrm{p}_1$: it can either be part of the set of three particles, or of the set of two. For this reason we have two different amplitudes and energy-momentum conserving Dirac deltas: one corresponding to the scattering $\mathbf{p}_1 + \mathbf{p}_2 + \mathbf{p}_3 \leftrightarrow \mathbf{p}_4 + \mathbf{p}_5$, and another to $\mathbf{p}_4 + \mathbf{p}_2 + \mathbf{p}_3 \leftrightarrow \mathbf{p}_1 + \mathbf{p}_5$. Once again, each of these two options has also its reversed version, which translates into a ``gain'' or a ``loss'' of a particle of momentum $\mathbf{p}_1$. Finally, note that the symmetry factors are different depending on whether the momentum $\mathbf{p}_1$ is part of the set of three or the set of two.

\subsubsection*{Continuity and Euler equations}

We now have the necessary ingredients to obtain the equations governing the evolution of the $\phi$-fluid: according to \Eqst{eq:A2}{eq:A5} we can multiply \Eq{eq:A6} by the appropriate functions of $\mathbf{p}_1$ and integrate to obtain the equations for $\rho$, $\delta$, $\theta$, and so forth. As we will show in this Appendix, for $\rho$, $\delta$, and $\theta$ the collision terms vanish, since their corresponding weights inside the integrals in \Eqs{eq:A3}{eq:A5} are the energy $E$ and momentum $\mathbf{p}$, which are conserved by the interactions.

From \Eq{eq:A6} we can obtain the evolution of $\rho$, $\delta$, and $\theta$ using \Eqs{eq:A3}{eq:A5}, to zeroth or first order in $\Psi$. The left hand side gives the well known results (\cite{Ma:1995ey}):
\bea
	&& \dot \rho + 3 \mH \bl( \rho + P \br) \ , \label{eq:A11} \\
	&& \dot \rho \delta + \rho \dot \delta + 3 \mH \bl( \rho \delta + \delta P \br) -  3 \dot \varphi (\rho + P) + (\rho + P) \theta \ , \label{eq:A12} \\
	&& (\rho + P) \dot\theta + \bl( \dot \rho + \dot P + 4 \mH(\rho + P) \br)\theta - k^2 (\rho + P)\psi - k^2\delta P + k^2 (\rho + P) \sigma \ . \label{eq:A13}
\eea
Here $(\rho + P)\sigma \equiv - \int \text{D}p \,\, \frac{p^2}{E} \bl( (\hat{k} \cdot \hat{p})^2 - \frac{1}{3} \br) \f0 \Psi$ is the anisotropic stress.

Let us now focus on the right hand side of \Eq{eq:A6}, the collision operators. Starting with the \twoto term we obtain:
\beq\label{eq:A14}
	\int \text{D}p_1 \,\, \frac{1}{2E_1} C_{22}[f_1]\, W_1 = \int \prod_{i=1}^{4} \mathrm{d}\Pi_i \,\,  F_{22}\, W_1 \ ,
\eeq
where $W_1 \equiv W(\mathbf{p}_1)$ is some weight function of the momentum $\mathbf{p}_1$, according to \Eqst{eq:A2}{eq:A5}: $W_1=1$ if we want the equations for $n$, $W_1=E_1$ for $\rho$ and $\delta$, $W_1=i \mathbf{k}\cdot\mathbf{p}_1$ for $\theta$, and so forth.

Since we are integrating over all the momenta we are free to relabel them at will. Changing $12 \leftrightarrow 34$ in the second term of \Eq{eq:A9} and making use of $\vert \mathcal{M}_{22}(p_3,p_4;p_1,p_2) \vert = \vert \mathcal{M}_{22}(p_1,p_2;p_3,p_4) \vert$ takes \Eq{eq:A14} to:
\beq\label{eq:A15}
\int \prod_{i=1}^{4} \mathrm{d}\Pi_i \,\, \vert \mathcal{M}_{22}(p_1,p_2;p_3,p_4) \vert^2 (2 \pi)^4 \delta^4 \bl( \Sigma_i p_i \br) \frac{1}{1!2!}f_1 f_2 \bl( -W_1 + W_3 \br) \ .
\eeq

Similarly the integrals remain the same if we exchange $1 \leftrightarrow 2$, and $3 \leftrightarrow 4$. Doing this and taking the half sum of these exchanges gives (\cite{Pitaevskii}):
\beq\label{eq:A16}
	\int \prod_{i=1}^{4} \mathrm{d}\Pi_i \,\, \vert  \mathcal{M}_{22}(p_1,p_2;p_3,p_4) \vert^2 (2 \pi)^4 \delta^4 \bl( \Sigma_i p_i \br) \frac{1}{2!2!} f_1 f_2\bl( -W_1 -W_2 + W_3 + W_4 \br) \ .
\eeq
It is clear that if the weight $W$ is a quantity conserved by the \twoto interactions (such as energy, momentum, or particle number) this collision term vanishes.

Let us now focus on the \threeto collision term:
\beq\label{eq:A17}
	\int \text{D}p_1 \,\, \frac{1}{2E_1} C_{32}[f_1] W_1 = \int \prod_{i=1}^{5} \mathrm{d}\Pi_i \,\, W_1 F_{32} \ .
\eeq

In the first term of $F_{32}$ (\Eq{eq:A10}) we can see that the labels 123 can be permuted without altering the result of the integration, while in the second term it is the 15 labels. Doing this to \Eq{eq:A17} yields:
\bea\label{eq:A18}
	\int \prod_{i=1}^{5} \mathrm{d}\Pi_i \,\, \frac{1}{3!2!} && \\
 && \!\!\! \!\!\! \!\!\! \!\!\! \!\!\! \!\!\! \!\!\! \!\!\! \!\!\! \!\!\! \bl[  \vert \mathcal{M}_{32}(p_1,p_2,p_3;p_4,p_5) \vert^2 (2 \pi)^4 \delta^4\bl( p_1\!+\!p_2 \!+\! p_3\!-\! p_4\!-\! p_5 \br)  \bl( -f_1 f_2 f_3 + f_4 f_5 \br) \bl( W_1\! +\! W_2 \!+\! W_3 \br) \br. \nonumber\\
	  && \!\!\!\!\!\! \!\!\!\!\!\!\!\!\!\!\!\! \!\!\! \!\!\!\!\! +\ \bl.  \vert \mathcal{M}_{32}(p_4,p_2,p_3;p_1,p_5) \vert^2 (2 \pi)^4 \delta^4 \bl( p_4\!+\!p_2 \!+\! p_3\!-\! p_1\!-\! p_5 \br) \bl( f_4 f_2 f_3 - f_1 f_5 \br) \bl( W_1\! +\! W_5 \br) \br] \!. \nonumber
\eea

Changing the labels $1 \leftrightarrow 4$ in the second term:
\bea\label{eq:A19}
	\int \prod_{i=1}^{5} \mathrm{d}\Pi_i \,\, \frac{1}{3!2!} \!\!\! && \!\!\! \vert \mathcal{M}_{32}(p_1,p_2,p_3;p_4,p_5) \vert^2 (2 \pi)^4 \delta^4\bl( p_1\!+\!p_2 \!+\! p_3\!-\! p_4\!-\! p_5 \br) \nonumber\\
	&\times& \!\!\! \bl( -f_1 f_2 f_3 + f_4 f_5 \br) \bl( W_1+ W_2 + W_3 - W_4 - W_5 \br) \ ,
\eea
which again vanishes if the weight $W$ is conserved by the \threeto collisions, like energy or momentum. Note that particle number is not conserved in these interactions. Combining the results for both collision terms with the left hands sides in \Eqst{eq:A11}{eq:A13} we get the standard equations for the background and perturbations of the $\phi$-fluid:
\bea
	\dot \rho + 3 \mH \bl( \rho + P \br) & = & 0 \ , \label{eq:A20} \\
	\dot\delta +(1+w) \bl( \theta - 3 \dot\varphi \br) + 3 \mH \bl( c_s^2 - w \br)\delta & = & 0 \ , \label{eq:A21}\\
	\dot\theta + \mH \bl( 1-3 c_s^2 \br)\theta - k^2 \bl( \psi + \frac{c_s^2}{1+w}\delta - \sigma \br) & = & 0 \ ; \label{eq:A22}
\eea
which are the continuity equation (for both background and perturbation energy densities) and the Euler equation. We have used the equation of state $w \equiv P/\rho$, and the fact that $\delta P = c_s^2 \rho \delta$ and $c_s^2 = \frac{\dot P}{\dot \rho}$ is the sound speed squared for adiabatic perturbations. Clearly from \Eq{eq:A20} $c_s^2 = w - \frac{\dot w}{3 \mH (1+w)}$.

Finally there remains the matter of the higher moments of the Boltzmann equation, which are obtained from \Eq{eq:A6} by performing integrations with the appropriate weights $W(\mathbf{p}_1)$, \eg the equation for $\dot \sigma$. It can be shown (\cite{Cyr-Racine:2015ihg}) that for fluids with very fast self-interactions (\ie perfect fluids) all these moments vanish. Taking our cannibal fluid to be one such fluid, with fast \twoto interactions, $\sigma = 0$ and then \Eqs{eq:A21}{eq:A22} reduce to \Eqs{eq:cannperts1a}{eq:cannperts1b}, and they are enough to describe the $\phi$ perturbations.

\subsubsection*{Number density and temperature}

We now write down the equation for the background number density $n$ and, together with \Eq{eq:A20} show that we can solve for the temperature as a function of the scale factor, $T(a)$. In order to obtain the equation for $n$ we integrate the Boltzmann equation \Eq{eq:A6} according to \Eq{eq:A2}, to zeroth order in $\Psi$. This corresponds to taking $W_1 = 1$ in \Eqs{eq:A16}{eq:A19} for the collision operators. The contribution from the operator $C_{22}$ vanishes, while that from $C_{32}$ does not, because \twoto interactions conserve particles number but \threeto do not. The result is:
\beq\label{eq:A23}
	\dot n + 3 \mH n = a \int \prod_{i=1}^{5} \mathrm{d}\Pi_i \,\, \frac{1}{3!2!} \vert \overline{\mathcal{M}_{32}} \vert^2 (2 \pi)^4 \delta^4 \bl( \Sigma_i \, p_i \br) \bl( -\f0_1 \f0_2 \f0_3 + \f0_4 \f0_5 \br)
\eeq

The right hand side is difficult to compute for general $\f0$. Nevertheless, we can make some simplifying assumptions. In particular, if we assume very fast \twoto self-interactions then the $\phi$-fluid is in thermal equilibrium, \ie there is a sensible temperature $T$ that can be associated with it. Therefore, under this assumption we can write:
\beq\label{eq:A24}
  f^{(0)} (p) = \frac{1}{e^{\frac{E-\mu}{T}}-1}
\eeq
for our bosonic cannibal dark matter particles, with $\mu$ their chemical potential.

We can further simplify the right-hand-side of the number density equation by concerning ourselves only with non-relativistic $\phi$ particles, since we know that when they are relativistic they are in chemical equilibrium. Indeed, in the relativistic regime dimensional analysis implies that the \threeto interactions rate is $\Gamma_{32} \sim \alpha^3 T \propto a^{-1}$, which remains bigger than the Hubble expansion rate of the Universe $H$ during both radiation and matter domination. Therefore for the relativistic case $\Gamma_{32} \gg H$ and the fluid is in chemical equilibrium where the number and energy densities can be determined
from entropy conservation. 

For non-relativistic particles $E \sim m \gg T$ and therefore we can write:
\beq\label{eq:A25}
  \f0(p) \simeq e^{\mu/T} \f0_\cheq \ , \quad \text{with } \quad \f0_\cheq \equiv e^{-E/T} \ ;
\eeq
the subindex standing for chemical equilibrium, when $\mu=0$. This means that we can write the $\phi$-fluid background quantities in terms of their values in chemical equilibrium:
\bea
  n \!\!& \simeq &\!\! e^{\mu/T} n_\cheq \ , \quad \rho \simeq e^{\mu/T} \rho_\cheq \ , \quad P \simeq e^{\mu/T} P_\cheq \ ; \label{eq:A26} \\
  \text{with} \quad n_\cheq \!\!& \approx &\!\! m^3\bl( \frac{T/m}{2 \pi} \br)^{3/2} e^{-m/T} \bl( 1 + \frac{15}{8}\frac{T}{m} + \frac{105}{128} \frac{T^2}{m^2} \br) \ , \label{eq:A27}\\ 
  \rho_\cheq \!\!& \approx &\!\! m^4\bl( \frac{T/m}{2 \pi} \br)^{3/2} e^{-m/T} \bl( 1 + \frac{27}{8}\frac{T}{m} + \frac{705}{128} \frac{T^2}{m^2} \br) \ , \label{eq:A28}\\
  P_\cheq \!\!& \approx &\!\! m^4\bl( \frac{T/m}{2 \pi} \br)^{3/2} e^{-m/T} \bl( \frac{T}{m} + \frac{15}{8}\frac{T^2}{m^2} \br) \ ; \label{eq:A29}
\eea
obtained simply by integrating \Eqst{eq:A2}{eq:A4} and expanding for $T/m \ll 1$. Note that $\rho_\cheq \approx m n_\cheq$ and $P_\cheq \approx T n_\cheq$ are valid to lowest order in $T/m$ and therefore only suitable for a qualitative analysis such as the one presented in the main body of this paper. In order to accurately solve the differential equations for $\rho$ and $n$ we use the expressions found in \Eqst{eq:A27}{eq:A29}.

With this in mind, we can finally write the background number density equation:
\bea
  \dot n + 3 \mH n & = & - a \langle \sigma_{32} v^2 \rangle n^2 \bl( n - n_\cheq \br) \ , \label{eq:A30} \\
  \langle \sigma_{32} v^2 \rangle & \equiv & \frac{1}{n_\cheq^3} \int \prod_{i=1}^{5} \mathrm{d}\Pi_i \,\, \frac{1}{3!2!} \ \f0_{\cheq,\, 1}\f0_{\cheq,\, 2}\f0_{\cheq,\, 3} \vert \mathcal{M}_{32}(p_i) \vert^2 (2 \pi)^4 \delta^4 \bl( \Sigma_i \, p_i \br) \ , \label{eq:A31}
\eea
where we have used $\f0_{\cheq,\, 4}\f0_{\cheq,\, 5} \approx e^{-\frac{E_4 + E_5}{T}} = e^{- \frac{E_1 + E_2 + E_3}{T}} \approx \f0_{\cheq,\, 1} \f0_{\cheq,\, 2} \f0_{\cheq,\, 3}$. As a consistency check, if $\Gamma_{32} \equiv \langle \sigma_{32} v^2 \rangle n^2 \gg H$ then $n \approx n_\eq$ and the dark matter is in chemical equilibrium (\ie $\mu \approx 0$).

With the aid of \Eq{eq:A26} we can write \Eq{eq:A20} also in terms of $\rho_\cheq$ and $e^{\mu/T}$. Doing this, and changing variables to the scale factor, \Eqs{eq:A20}{eq:A30} become:
\bea
	a \frac{d }{d a} \bl( e^{\mu/T} \rho_\cheq \br) + 3 e^{\mu/T} \bl( \rho_\cheq + P_\cheq \br) & = & 0 \ ,  \nonumber\\
	a \frac{d}{d a} \bl( e^{\mu/T} n_\cheq \br) + 3 e^{\mu/T} n_\cheq & = & - \frac{\langle \sigma_{32} v^2 \rangle n_\cheq^3}{H} \bl( e^{3 \mu/T} - e^{2 \mu/T} \br) \ . \label{eq:A32}
\eea

These are two coupled differential equations for $\mu(a)$ and $T(a)$, which we solve numerically in order to obtain the evolution of the background quantities of the $\phi$-fluid.

\bibliography{can}{}

\begin{thebibliography}{61}
\expandafter\ifx\csname natexlab\endcsname\relax\def\natexlab#1{#1}\fi
\expandafter\ifx\csname bibnamefont\endcsname\relax
  \def\bibnamefont#1{#1}\fi
\expandafter\ifx\csname bibfnamefont\endcsname\relax
  \def\bibfnamefont#1{#1}\fi
\expandafter\ifx\csname citenamefont\endcsname\relax
  \def\citenamefont#1{#1}\fi
\expandafter\ifx\csname url\endcsname\relax
  \def\url#1{\texttt{#1}}\fi
\expandafter\ifx\csname urlprefix\endcsname\relax\def\urlprefix{URL }\fi
\providecommand{\bibinfo}[2]{#2}
\providecommand{\eprint}[2][]{\url{#2}}

\bibitem[{\citenamefont{Riess et~al.}(2016)}]{Riess:2016jrr}
\bibinfo{author}{\bibfnamefont{A.~G.} \bibnamefont{Riess}}
  \bibnamefont{et~al.}, \bibinfo{journal}{Astrophys. J.}
  \textbf{\bibinfo{volume}{826}}, \bibinfo{pages}{56} (\bibinfo{year}{2016}),
  \eprint{1604.01424}.

\bibitem[{\citenamefont{Bonvin et~al.}(2017)}]{Bonvin:2016crt}
\bibinfo{author}{\bibfnamefont{V.}~\bibnamefont{Bonvin}} \bibnamefont{et~al.},
  \bibinfo{journal}{Mon. Not. Roy. Astron. Soc.}
  \textbf{\bibinfo{volume}{465}}, \bibinfo{pages}{4914} (\bibinfo{year}{2017}),
  \eprint{1607.01790}.

\bibitem[{\citenamefont{Heymans et~al.}(2013)}]{Heymans:2013fya}
\bibinfo{author}{\bibfnamefont{C.}~\bibnamefont{Heymans}} \bibnamefont{et~al.},
  \bibinfo{journal}{Mon. Not. Roy. Astron. Soc.}
  \textbf{\bibinfo{volume}{432}}, \bibinfo{pages}{2433} (\bibinfo{year}{2013}),
  \eprint{1303.1808}.

\bibitem[{\citenamefont{Joudaki et~al.}(2017{\natexlab{a}})}]{Joudaki:2016mvz}
\bibinfo{author}{\bibfnamefont{S.}~\bibnamefont{Joudaki}} \bibnamefont{et~al.},
  \bibinfo{journal}{Mon. Not. Roy. Astron. Soc.}
  \textbf{\bibinfo{volume}{465}}, \bibinfo{pages}{2033}
  (\bibinfo{year}{2017}{\natexlab{a}}), \eprint{1601.05786}.

\bibitem[{\citenamefont{Ade et~al.}(2016)}]{Ade:2015fva}
\bibinfo{author}{\bibfnamefont{P.~A.~R.} \bibnamefont{Ade}}
  \bibnamefont{et~al.} (\bibinfo{collaboration}{Planck}),
  \bibinfo{journal}{Astron. Astrophys.} \textbf{\bibinfo{volume}{594}},
  \bibinfo{pages}{A24} (\bibinfo{year}{2016}), \eprint{1502.01597}.

\bibitem[{\citenamefont{Ade et~al.}(2014)}]{Ade:2013lmv}
\bibinfo{author}{\bibfnamefont{P.~A.~R.} \bibnamefont{Ade}}
  \bibnamefont{et~al.} (\bibinfo{collaboration}{Planck}),
  \bibinfo{journal}{Astron. Astrophys.} \textbf{\bibinfo{volume}{571}},
  \bibinfo{pages}{A20} (\bibinfo{year}{2014}), \eprint{1303.5080}.

\bibitem[{\citenamefont{Kohlinger et~al.}(2017)}]{Kohlinger:2017sxk}
\bibinfo{author}{\bibfnamefont{F.}~\bibnamefont{Kohlinger}}
  \bibnamefont{et~al.} (\bibinfo{year}{2017}), \eprint{1706.02892}.

\bibitem[{\citenamefont{Joudaki et~al.}(2017{\natexlab{b}})}]{Joudaki:2017zdt}
\bibinfo{author}{\bibfnamefont{S.}~\bibnamefont{Joudaki}} \bibnamefont{et~al.}
  (\bibinfo{year}{2017}{\natexlab{b}}), \eprint{1707.06627}.

\bibitem[{\citenamefont{Buen-Abad et~al.}(2015)\citenamefont{Buen-Abad,
  Marques-Tavares, and Schmaltz}}]{Buen-Abad:2015ova}
\bibinfo{author}{\bibfnamefont{M.~A.} \bibnamefont{Buen-Abad}},
  \bibinfo{author}{\bibfnamefont{G.}~\bibnamefont{Marques-Tavares}},
  \bibnamefont{and} \bibinfo{author}{\bibfnamefont{M.}~\bibnamefont{Schmaltz}},
  \bibinfo{journal}{Phys. Rev.} \textbf{\bibinfo{volume}{D92}},
  \bibinfo{pages}{023531} (\bibinfo{year}{2015}), \eprint{1505.03542}.

\bibitem[{\citenamefont{Lesgourgues et~al.}(2016)\citenamefont{Lesgourgues,
  Marques-Tavares, and Schmaltz}}]{Lesgourgues:2015wza}
\bibinfo{author}{\bibfnamefont{J.}~\bibnamefont{Lesgourgues}},
  \bibinfo{author}{\bibfnamefont{G.}~\bibnamefont{Marques-Tavares}},
  \bibnamefont{and} \bibinfo{author}{\bibfnamefont{M.}~\bibnamefont{Schmaltz}},
  \bibinfo{journal}{JCAP} \textbf{\bibinfo{volume}{1602}}, \bibinfo{pages}{037}
  (\bibinfo{year}{2016}), \eprint{1507.04351}.

\bibitem[{\citenamefont{Chacko et~al.}(2016)\citenamefont{Chacko, Cui, Hong,
  Okui, and Tsai}}]{Chacko:2016kgg}
\bibinfo{author}{\bibfnamefont{Z.}~\bibnamefont{Chacko}},
  \bibinfo{author}{\bibfnamefont{Y.}~\bibnamefont{Cui}},
  \bibinfo{author}{\bibfnamefont{S.}~\bibnamefont{Hong}},
  \bibinfo{author}{\bibfnamefont{T.}~\bibnamefont{Okui}}, \bibnamefont{and}
  \bibinfo{author}{\bibfnamefont{Y.}~\bibnamefont{Tsai}},
  \bibinfo{journal}{JHEP} \textbf{\bibinfo{volume}{12}}, \bibinfo{pages}{108}
  (\bibinfo{year}{2016}), \eprint{1609.03569}.

\bibitem[{\citenamefont{Poulin et~al.}(2016)\citenamefont{Poulin, Serpico, and
  Lesgourgues}}]{Poulin:2016nat}
\bibinfo{author}{\bibfnamefont{V.}~\bibnamefont{Poulin}},
  \bibinfo{author}{\bibfnamefont{P.~D.} \bibnamefont{Serpico}},
  \bibnamefont{and}
  \bibinfo{author}{\bibfnamefont{J.}~\bibnamefont{Lesgourgues}},
  \bibinfo{journal}{JCAP} \textbf{\bibinfo{volume}{1608}}, \bibinfo{pages}{036}
  (\bibinfo{year}{2016}), \eprint{1606.02073}.

\bibitem[{\citenamefont{MacCrann et~al.}(2015)\citenamefont{MacCrann, Zuntz,
  Bridle, Jain, and Becker}}]{MacCrann:2014wfa}
\bibinfo{author}{\bibfnamefont{N.}~\bibnamefont{MacCrann}},
  \bibinfo{author}{\bibfnamefont{J.}~\bibnamefont{Zuntz}},
  \bibinfo{author}{\bibfnamefont{S.}~\bibnamefont{Bridle}},
  \bibinfo{author}{\bibfnamefont{B.}~\bibnamefont{Jain}}, \bibnamefont{and}
  \bibinfo{author}{\bibfnamefont{M.~R.} \bibnamefont{Becker}},
  \bibinfo{journal}{Mon. Not. Roy. Astron. Soc.}
  \textbf{\bibinfo{volume}{451}}, \bibinfo{pages}{2877} (\bibinfo{year}{2015}),
  \eprint{1408.4742}.

\bibitem[{\citenamefont{Canac et~al.}(2016)\citenamefont{Canac, Aslanyan,
  Abazajian, Easther, and Price}}]{Canac:2016smv}
\bibinfo{author}{\bibfnamefont{N.}~\bibnamefont{Canac}},
  \bibinfo{author}{\bibfnamefont{G.}~\bibnamefont{Aslanyan}},
  \bibinfo{author}{\bibfnamefont{K.~N.} \bibnamefont{Abazajian}},
  \bibinfo{author}{\bibfnamefont{R.}~\bibnamefont{Easther}}, \bibnamefont{and}
  \bibinfo{author}{\bibfnamefont{L.~C.} \bibnamefont{Price}},
  \bibinfo{journal}{JCAP} \textbf{\bibinfo{volume}{1609}}, \bibinfo{pages}{022}
  (\bibinfo{year}{2016}), \eprint{1606.03057}.

\bibitem[{\citenamefont{Bernal et~al.}(2016{\natexlab{a}})\citenamefont{Bernal,
  Verde, and Riess}}]{Bernal:2016gxb}
\bibinfo{author}{\bibfnamefont{J.~L.} \bibnamefont{Bernal}},
  \bibinfo{author}{\bibfnamefont{L.}~\bibnamefont{Verde}}, \bibnamefont{and}
  \bibinfo{author}{\bibfnamefont{A.~G.} \bibnamefont{Riess}},
  \bibinfo{journal}{JCAP} \textbf{\bibinfo{volume}{1610}}, \bibinfo{pages}{019}
  (\bibinfo{year}{2016}{\natexlab{a}}), \eprint{1607.05617}.

\bibitem[{\citenamefont{Chudaykin et~al.}(2016)\citenamefont{Chudaykin,
  Gorbunov, and Tkachev}}]{Chudaykin:2016yfk}
\bibinfo{author}{\bibfnamefont{A.}~\bibnamefont{Chudaykin}},
  \bibinfo{author}{\bibfnamefont{D.}~\bibnamefont{Gorbunov}}, \bibnamefont{and}
  \bibinfo{author}{\bibfnamefont{I.}~\bibnamefont{Tkachev}},
  \bibinfo{journal}{Phys. Rev.} \textbf{\bibinfo{volume}{D94}},
  \bibinfo{pages}{023528} (\bibinfo{year}{2016}), \eprint{1602.08121}.

\bibitem[{\citenamefont{Archidiacono et~al.}(2016)\citenamefont{Archidiacono,
  Gariazzo, Giunti, Hannestad, Hansen, Laveder, and
  Tram}}]{Archidiacono:2016kkh}
\bibinfo{author}{\bibfnamefont{M.}~\bibnamefont{Archidiacono}},
  \bibinfo{author}{\bibfnamefont{S.}~\bibnamefont{Gariazzo}},
  \bibinfo{author}{\bibfnamefont{C.}~\bibnamefont{Giunti}},
  \bibinfo{author}{\bibfnamefont{S.}~\bibnamefont{Hannestad}},
  \bibinfo{author}{\bibfnamefont{R.}~\bibnamefont{Hansen}},
  \bibinfo{author}{\bibfnamefont{M.}~\bibnamefont{Laveder}}, \bibnamefont{and}
  \bibinfo{author}{\bibfnamefont{T.}~\bibnamefont{Tram}},
  \bibinfo{journal}{JCAP} \textbf{\bibinfo{volume}{1608}}, \bibinfo{pages}{067}
  (\bibinfo{year}{2016}), \eprint{1606.07673}.

\bibitem[{\citenamefont{Joudaki et~al.}(2017{\natexlab{c}})}]{Joudaki:2016kym}
\bibinfo{author}{\bibfnamefont{S.}~\bibnamefont{Joudaki}} \bibnamefont{et~al.},
  \bibinfo{journal}{Mon. Not. Roy. Astron. Soc.}
  \textbf{\bibinfo{volume}{471}}, \bibinfo{pages}{1259}
  (\bibinfo{year}{2017}{\natexlab{c}}), \eprint{1610.04606}.

\bibitem[{\citenamefont{Buen-Abad et~al.}(2018)\citenamefont{Buen-Abad,
  Schmaltz, Lesgourgues, and Brinckmann}}]{Buen-Abad:2017gxg}
\bibinfo{author}{\bibfnamefont{M.~A.} \bibnamefont{Buen-Abad}},
  \bibinfo{author}{\bibfnamefont{M.}~\bibnamefont{Schmaltz}},
  \bibinfo{author}{\bibfnamefont{J.}~\bibnamefont{Lesgourgues}},
  \bibnamefont{and}
  \bibinfo{author}{\bibfnamefont{T.}~\bibnamefont{Brinckmann}},
  \bibinfo{journal}{JCAP} \textbf{\bibinfo{volume}{1801}}, \bibinfo{pages}{008}
  (\bibinfo{year}{2018}), \eprint{1708.09406}.

\bibitem[{\citenamefont{Raveri et~al.}(2017)\citenamefont{Raveri, Hu, Hoffman,
  and Wang}}]{Raveri:2017jto}
\bibinfo{author}{\bibfnamefont{M.}~\bibnamefont{Raveri}},
  \bibinfo{author}{\bibfnamefont{W.}~\bibnamefont{Hu}},
  \bibinfo{author}{\bibfnamefont{T.}~\bibnamefont{Hoffman}}, \bibnamefont{and}
  \bibinfo{author}{\bibfnamefont{L.-T.} \bibnamefont{Wang}},
  \bibinfo{journal}{Phys. Rev.} \textbf{\bibinfo{volume}{D96}},
  \bibinfo{pages}{103501} (\bibinfo{year}{2017}), \eprint{1709.04877}.

\bibitem[{\citenamefont{Lancaster et~al.}(2017)\citenamefont{Lancaster,
  Cyr-Racine, Knox, and Pan}}]{Lancaster:2017ksf}
\bibinfo{author}{\bibfnamefont{L.}~\bibnamefont{Lancaster}},
  \bibinfo{author}{\bibfnamefont{F.-Y.} \bibnamefont{Cyr-Racine}},
  \bibinfo{author}{\bibfnamefont{L.}~\bibnamefont{Knox}}, \bibnamefont{and}
  \bibinfo{author}{\bibfnamefont{Z.}~\bibnamefont{Pan}},
  \bibinfo{journal}{JCAP} \textbf{\bibinfo{volume}{1707}}, \bibinfo{pages}{033}
  (\bibinfo{year}{2017}), \eprint{1704.06657}.

\bibitem[{\citenamefont{Oldengott et~al.}(2017)\citenamefont{Oldengott, Tram,
  Rampf, and Wong}}]{Oldengott:2017fhy}
\bibinfo{author}{\bibfnamefont{I.~M.} \bibnamefont{Oldengott}},
  \bibinfo{author}{\bibfnamefont{T.}~\bibnamefont{Tram}},
  \bibinfo{author}{\bibfnamefont{C.}~\bibnamefont{Rampf}}, \bibnamefont{and}
  \bibinfo{author}{\bibfnamefont{Y.~Y.~Y.} \bibnamefont{Wong}}
  (\bibinfo{year}{2017}), \eprint{1706.02123}.

\bibitem[{\citenamefont{Ko and Tang}(2017)}]{Ko:2016fcd}
\bibinfo{author}{\bibfnamefont{P.}~\bibnamefont{Ko}} \bibnamefont{and}
  \bibinfo{author}{\bibfnamefont{Y.}~\bibnamefont{Tang}},
  \bibinfo{journal}{Phys. Lett.} \textbf{\bibinfo{volume}{B768}},
  \bibinfo{pages}{12} (\bibinfo{year}{2017}), \eprint{1609.02307}.

\bibitem[{\citenamefont{Ko and Tang}(2016)}]{Ko:2016uft}
\bibinfo{author}{\bibfnamefont{P.}~\bibnamefont{Ko}} \bibnamefont{and}
  \bibinfo{author}{\bibfnamefont{Y.}~\bibnamefont{Tang}},
  \bibinfo{journal}{Phys. Lett.} \textbf{\bibinfo{volume}{B762}},
  \bibinfo{pages}{462} (\bibinfo{year}{2016}), \eprint{1608.01083}.

\bibitem[{\citenamefont{Ko et~al.}(2017)\citenamefont{Ko, Nagata, and
  Tang}}]{Ko:2017uyb}
\bibinfo{author}{\bibfnamefont{P.}~\bibnamefont{Ko}},
  \bibinfo{author}{\bibfnamefont{N.}~\bibnamefont{Nagata}}, \bibnamefont{and}
  \bibinfo{author}{\bibfnamefont{Y.}~\bibnamefont{Tang}}
  (\bibinfo{year}{2017}), \eprint{1706.05605}.

\bibitem[{\citenamefont{Chacko et~al.}(2018)\citenamefont{Chacko, Curtin,
  Geller, and Tsai}}]{Chacko:2018vss}
\bibinfo{author}{\bibfnamefont{Z.}~\bibnamefont{Chacko}},
  \bibinfo{author}{\bibfnamefont{D.}~\bibnamefont{Curtin}},
  \bibinfo{author}{\bibfnamefont{M.}~\bibnamefont{Geller}}, \bibnamefont{and}
  \bibinfo{author}{\bibfnamefont{Y.}~\bibnamefont{Tsai}}
  (\bibinfo{year}{2018}), \eprint{1803.03263}.

\bibitem[{\citenamefont{Poulin et~al.}(2018)\citenamefont{Poulin, Boddy, Bird,
  and Kamionkowski}}]{Poulin:2018zxs}
\bibinfo{author}{\bibfnamefont{V.}~\bibnamefont{Poulin}},
  \bibinfo{author}{\bibfnamefont{K.~K.} \bibnamefont{Boddy}},
  \bibinfo{author}{\bibfnamefont{S.}~\bibnamefont{Bird}}, \bibnamefont{and}
  \bibinfo{author}{\bibfnamefont{M.}~\bibnamefont{Kamionkowski}}
  (\bibinfo{year}{2018}), \eprint{1803.02474}.

\bibitem[{\citenamefont{Pan et~al.}(2018)\citenamefont{Pan, Kaplinghat, and
  Knox}}]{Pan:2018zha}
\bibinfo{author}{\bibfnamefont{Z.}~\bibnamefont{Pan}},
  \bibinfo{author}{\bibfnamefont{M.}~\bibnamefont{Kaplinghat}},
  \bibnamefont{and} \bibinfo{author}{\bibfnamefont{L.}~\bibnamefont{Knox}}
  (\bibinfo{year}{2018}), \eprint{1801.07348}.

\bibitem[{\citenamefont{Carlson et~al.}(1992)\citenamefont{Carlson, Machacek,
  and Hall}}]{Carlson:1992fn}
\bibinfo{author}{\bibfnamefont{E.~D.} \bibnamefont{Carlson}},
  \bibinfo{author}{\bibfnamefont{M.~E.} \bibnamefont{Machacek}},
  \bibnamefont{and} \bibinfo{author}{\bibfnamefont{L.~J.} \bibnamefont{Hall}},
  \bibinfo{journal}{Astrophys. J.} \textbf{\bibinfo{volume}{398}},
  \bibinfo{pages}{43} (\bibinfo{year}{1992}).

\bibitem[{\citenamefont{Machacek}(1994)}]{Machacek:1994vg}
\bibinfo{author}{\bibfnamefont{M.~E.} \bibnamefont{Machacek}},
  \bibinfo{journal}{Astrophys. J.} \textbf{\bibinfo{volume}{431}},
  \bibinfo{pages}{41} (\bibinfo{year}{1994}).

\bibitem[{\citenamefont{de~Laix et~al.}(1995)\citenamefont{de~Laix, Scherrer,
  and Schaefer}}]{deLaix:1995vi}
\bibinfo{author}{\bibfnamefont{A.~A.} \bibnamefont{de~Laix}},
  \bibinfo{author}{\bibfnamefont{R.~J.} \bibnamefont{Scherrer}},
  \bibnamefont{and} \bibinfo{author}{\bibfnamefont{R.~K.}
  \bibnamefont{Schaefer}}, \bibinfo{journal}{Astrophys. J.}
  \textbf{\bibinfo{volume}{452}}, \bibinfo{pages}{495} (\bibinfo{year}{1995}),
  \eprint{astro-ph/9502087}.

\bibitem[{\citenamefont{Soni and Zhang}(2016)}]{Soni:2016gzf}
\bibinfo{author}{\bibfnamefont{A.}~\bibnamefont{Soni}} \bibnamefont{and}
  \bibinfo{author}{\bibfnamefont{Y.}~\bibnamefont{Zhang}},
  \bibinfo{journal}{Phys. Rev.} \textbf{\bibinfo{volume}{D93}},
  \bibinfo{pages}{115025} (\bibinfo{year}{2016}), \eprint{1602.00714}.

\bibitem[{\citenamefont{Pappadopulo et~al.}(2016)\citenamefont{Pappadopulo,
  Ruderman, and Trevisan}}]{Pappadopulo:2016pkp}
\bibinfo{author}{\bibfnamefont{D.}~\bibnamefont{Pappadopulo}},
  \bibinfo{author}{\bibfnamefont{J.~T.} \bibnamefont{Ruderman}},
  \bibnamefont{and} \bibinfo{author}{\bibfnamefont{G.}~\bibnamefont{Trevisan}},
  \bibinfo{journal}{Phys. Rev.} \textbf{\bibinfo{volume}{D94}},
  \bibinfo{pages}{035005} (\bibinfo{year}{2016}), \eprint{1602.04219}.

\bibitem[{\citenamefont{Dey et~al.}(2017)\citenamefont{Dey, Maity, and
  Ray}}]{Dey:2016qgf}
\bibinfo{author}{\bibfnamefont{U.~K.} \bibnamefont{Dey}},
  \bibinfo{author}{\bibfnamefont{T.~N.} \bibnamefont{Maity}}, \bibnamefont{and}
  \bibinfo{author}{\bibfnamefont{T.~S.} \bibnamefont{Ray}},
  \bibinfo{journal}{JCAP} \textbf{\bibinfo{volume}{1703}}, \bibinfo{pages}{045}
  (\bibinfo{year}{2017}), \eprint{1612.09074}.

\bibitem[{\citenamefont{Bernal et~al.}(2016{\natexlab{b}})\citenamefont{Bernal,
  Chu, Garcia-Cely, Hambye, and Zaldivar}}]{Bernal:2015ova}
\bibinfo{author}{\bibfnamefont{N.}~\bibnamefont{Bernal}},
  \bibinfo{author}{\bibfnamefont{X.}~\bibnamefont{Chu}},
  \bibinfo{author}{\bibfnamefont{C.}~\bibnamefont{Garcia-Cely}},
  \bibinfo{author}{\bibfnamefont{T.}~\bibnamefont{Hambye}}, \bibnamefont{and}
  \bibinfo{author}{\bibfnamefont{B.}~\bibnamefont{Zaldivar}},
  \bibinfo{journal}{JCAP} \textbf{\bibinfo{volume}{1603}}, \bibinfo{pages}{018}
  (\bibinfo{year}{2016}{\natexlab{b}}), \eprint{1510.08063}.

\bibitem[{\citenamefont{Kuflik et~al.}(2016)\citenamefont{Kuflik, Perelstein,
  Lorier, and Tsai}}]{Kuflik:2015isi}
\bibinfo{author}{\bibfnamefont{E.}~\bibnamefont{Kuflik}},
  \bibinfo{author}{\bibfnamefont{M.}~\bibnamefont{Perelstein}},
  \bibinfo{author}{\bibfnamefont{N.~R.-L.} \bibnamefont{Lorier}},
  \bibnamefont{and} \bibinfo{author}{\bibfnamefont{Y.-D.} \bibnamefont{Tsai}},
  \bibinfo{journal}{Phys. Rev. Lett.} \textbf{\bibinfo{volume}{116}},
  \bibinfo{pages}{221302} (\bibinfo{year}{2016}), \eprint{1512.04545}.

\bibitem[{\citenamefont{Hochberg et~al.}(2014)\citenamefont{Hochberg, Kuflik,
  Volansky, and Wacker}}]{Hochberg:2014dra}
\bibinfo{author}{\bibfnamefont{Y.}~\bibnamefont{Hochberg}},
  \bibinfo{author}{\bibfnamefont{E.}~\bibnamefont{Kuflik}},
  \bibinfo{author}{\bibfnamefont{T.}~\bibnamefont{Volansky}}, \bibnamefont{and}
  \bibinfo{author}{\bibfnamefont{J.~G.} \bibnamefont{Wacker}},
  \bibinfo{journal}{Phys. Rev. Lett.} \textbf{\bibinfo{volume}{113}},
  \bibinfo{pages}{171301} (\bibinfo{year}{2014}), \eprint{1402.5143}.

\bibitem[{\citenamefont{Hochberg et~al.}(2015)\citenamefont{Hochberg, Kuflik,
  Murayama, Volansky, and Wacker}}]{Hochberg:2014kqa}
\bibinfo{author}{\bibfnamefont{Y.}~\bibnamefont{Hochberg}},
  \bibinfo{author}{\bibfnamefont{E.}~\bibnamefont{Kuflik}},
  \bibinfo{author}{\bibfnamefont{H.}~\bibnamefont{Murayama}},
  \bibinfo{author}{\bibfnamefont{T.}~\bibnamefont{Volansky}}, \bibnamefont{and}
  \bibinfo{author}{\bibfnamefont{J.~G.} \bibnamefont{Wacker}},
  \bibinfo{journal}{Phys. Rev. Lett.} \textbf{\bibinfo{volume}{115}},
  \bibinfo{pages}{021301} (\bibinfo{year}{2015}), \eprint{1411.3727}.

\bibitem[{\citenamefont{Bernal et~al.}(2015)\citenamefont{Bernal, Garcia-Cely,
  and Rosenfeld}}]{Bernal:2015bla}
\bibinfo{author}{\bibfnamefont{N.}~\bibnamefont{Bernal}},
  \bibinfo{author}{\bibfnamefont{C.}~\bibnamefont{Garcia-Cely}},
  \bibnamefont{and}
  \bibinfo{author}{\bibfnamefont{R.}~\bibnamefont{Rosenfeld}},
  \bibinfo{journal}{JCAP} \textbf{\bibinfo{volume}{1504}}, \bibinfo{pages}{012}
  (\bibinfo{year}{2015}), \eprint{1501.01973}.

\bibitem[{\citenamefont{Bernal and Chu}(2016)}]{Bernal:2015xba}
\bibinfo{author}{\bibfnamefont{N.}~\bibnamefont{Bernal}} \bibnamefont{and}
  \bibinfo{author}{\bibfnamefont{X.}~\bibnamefont{Chu}},
  \bibinfo{journal}{JCAP} \textbf{\bibinfo{volume}{1601}}, \bibinfo{pages}{006}
  (\bibinfo{year}{2016}), \eprint{1510.08527}.

\bibitem[{\citenamefont{Bernal et~al.}(2017)\citenamefont{Bernal, Chu, and
  Pradler}}]{Bernal:2017mqb}
\bibinfo{author}{\bibfnamefont{N.}~\bibnamefont{Bernal}},
  \bibinfo{author}{\bibfnamefont{X.}~\bibnamefont{Chu}}, \bibnamefont{and}
  \bibinfo{author}{\bibfnamefont{J.}~\bibnamefont{Pradler}},
  \bibinfo{journal}{Phys. Rev.} \textbf{\bibinfo{volume}{D95}},
  \bibinfo{pages}{115023} (\bibinfo{year}{2017}), \eprint{1702.04906}.

\bibitem[{\citenamefont{Lee and Seo}(2015)}]{Lee:2015gsa}
\bibinfo{author}{\bibfnamefont{H.~M.} \bibnamefont{Lee}} \bibnamefont{and}
  \bibinfo{author}{\bibfnamefont{M.-S.} \bibnamefont{Seo}},
  \bibinfo{journal}{Phys. Lett.} \textbf{\bibinfo{volume}{B748}},
  \bibinfo{pages}{316} (\bibinfo{year}{2015}), \eprint{1504.00745}.

\bibitem[{\citenamefont{Kamada et~al.}(2016)\citenamefont{Kamada, Yamada,
  Yanagida, and Yonekura}}]{Kamada:2016ois}
\bibinfo{author}{\bibfnamefont{A.}~\bibnamefont{Kamada}},
  \bibinfo{author}{\bibfnamefont{M.}~\bibnamefont{Yamada}},
  \bibinfo{author}{\bibfnamefont{T.~T.} \bibnamefont{Yanagida}},
  \bibnamefont{and} \bibinfo{author}{\bibfnamefont{K.}~\bibnamefont{Yonekura}},
  \bibinfo{journal}{Phys. Rev.} \textbf{\bibinfo{volume}{D94}},
  \bibinfo{pages}{055035} (\bibinfo{year}{2016}), \eprint{1606.01628}.

\bibitem[{\citenamefont{Choi et~al.}(2017{\natexlab{a}})\citenamefont{Choi,
  Lee, and Seo}}]{Choi:2017mkk}
\bibinfo{author}{\bibfnamefont{S.-M.} \bibnamefont{Choi}},
  \bibinfo{author}{\bibfnamefont{H.~M.} \bibnamefont{Lee}}, \bibnamefont{and}
  \bibinfo{author}{\bibfnamefont{M.-S.} \bibnamefont{Seo}},
  \bibinfo{journal}{JHEP} \textbf{\bibinfo{volume}{04}}, \bibinfo{pages}{154}
  (\bibinfo{year}{2017}{\natexlab{a}}), \eprint{1702.07860}.

\bibitem[{\citenamefont{Choi et~al.}(2017{\natexlab{b}})\citenamefont{Choi,
  Hochberg, Kuflik, Lee, Mambrini, Murayama, and Pierre}}]{Choi:2017zww}
\bibinfo{author}{\bibfnamefont{S.-M.} \bibnamefont{Choi}},
  \bibinfo{author}{\bibfnamefont{Y.}~\bibnamefont{Hochberg}},
  \bibinfo{author}{\bibfnamefont{E.}~\bibnamefont{Kuflik}},
  \bibinfo{author}{\bibfnamefont{H.~M.} \bibnamefont{Lee}},
  \bibinfo{author}{\bibfnamefont{Y.}~\bibnamefont{Mambrini}},
  \bibinfo{author}{\bibfnamefont{H.}~\bibnamefont{Murayama}}, \bibnamefont{and}
  \bibinfo{author}{\bibfnamefont{M.}~\bibnamefont{Pierre}},
  \bibinfo{journal}{JHEP} \textbf{\bibinfo{volume}{10}}, \bibinfo{pages}{162}
  (\bibinfo{year}{2017}{\natexlab{b}}), \eprint{1707.01434}.

\bibitem[{\citenamefont{Choi et~al.}(2018)\citenamefont{Choi, Lee, Ko, and
  Natale}}]{Choi:2018iit}
\bibinfo{author}{\bibfnamefont{S.-M.} \bibnamefont{Choi}},
  \bibinfo{author}{\bibfnamefont{H.~M.} \bibnamefont{Lee}},
  \bibinfo{author}{\bibfnamefont{P.}~\bibnamefont{Ko}}, \bibnamefont{and}
  \bibinfo{author}{\bibfnamefont{A.}~\bibnamefont{Natale}}
  (\bibinfo{year}{2018}), \eprint{1801.07726}.

\bibitem[{\citenamefont{Boddy et~al.}(2014)\citenamefont{Boddy, Feng,
  Kaplinghat, Shadmi, and Tait}}]{Boddy:2014qxa}
\bibinfo{author}{\bibfnamefont{K.~K.} \bibnamefont{Boddy}},
  \bibinfo{author}{\bibfnamefont{J.~L.} \bibnamefont{Feng}},
  \bibinfo{author}{\bibfnamefont{M.}~\bibnamefont{Kaplinghat}},
  \bibinfo{author}{\bibfnamefont{Y.}~\bibnamefont{Shadmi}}, \bibnamefont{and}
  \bibinfo{author}{\bibfnamefont{T.~M.~P.} \bibnamefont{Tait}},
  \bibinfo{journal}{Phys. Rev.} \textbf{\bibinfo{volume}{D90}},
  \bibinfo{pages}{095016} (\bibinfo{year}{2014}), \eprint{1408.6532}.

\bibitem[{\citenamefont{Ma and Bertschinger}(1995)}]{Ma:1995ey}
\bibinfo{author}{\bibfnamefont{C.-P.} \bibnamefont{Ma}} \bibnamefont{and}
  \bibinfo{author}{\bibfnamefont{E.}~\bibnamefont{Bertschinger}},
  \bibinfo{journal}{Astrophys. J.} \textbf{\bibinfo{volume}{455}},
  \bibinfo{pages}{7} (\bibinfo{year}{1995}), \eprint{astro-ph/9506072}.

\bibitem[{\citenamefont{Steigman}(2012)}]{Steigman:2012ve}
\bibinfo{author}{\bibfnamefont{G.}~\bibnamefont{Steigman}},
  \bibinfo{journal}{Adv. High Energy Phys.} \textbf{\bibinfo{volume}{2012}},
  \bibinfo{pages}{268321} (\bibinfo{year}{2012}), \eprint{1208.0032}.

\bibitem[{\citenamefont{Lesgourgues et~al.}(2013)\citenamefont{Lesgourgues,
  Mangano, Miele, and Pastor}}]{Lesgourgues:1519137}
\bibinfo{author}{\bibfnamefont{J.}~\bibnamefont{Lesgourgues}},
  \bibinfo{author}{\bibfnamefont{G.}~\bibnamefont{Mangano}},
  \bibinfo{author}{\bibfnamefont{G.}~\bibnamefont{Miele}}, \bibnamefont{and}
  \bibinfo{author}{\bibfnamefont{S.}~\bibnamefont{Pastor}},
  \emph{\bibinfo{title}{{Neutrino cosmology}}} (\bibinfo{publisher}{Cambridge
  Univ. Press}, \bibinfo{address}{Cambridge}, \bibinfo{year}{2013}),
  \urlprefix\url{https://cds.cern.ch/record/1519137}.

\bibitem[{\citenamefont{Cornwall and Soni}(1983)}]{Cornwall:1982zn}
\bibinfo{author}{\bibfnamefont{J.~M.} \bibnamefont{Cornwall}} \bibnamefont{and}
  \bibinfo{author}{\bibfnamefont{A.}~\bibnamefont{Soni}},
  \bibinfo{journal}{Phys. Lett.} \textbf{\bibinfo{volume}{120B}},
  \bibinfo{pages}{431} (\bibinfo{year}{1983}).

\bibitem[{\citenamefont{Morningstar and Peardon}(1999)}]{Morningstar:1999rf}
\bibinfo{author}{\bibfnamefont{C.~J.} \bibnamefont{Morningstar}}
  \bibnamefont{and} \bibinfo{author}{\bibfnamefont{M.~J.}
  \bibnamefont{Peardon}}, \bibinfo{journal}{Phys. Rev.}
  \textbf{\bibinfo{volume}{D60}}, \bibinfo{pages}{034509}
  (\bibinfo{year}{1999}), \eprint{hep-lat/9901004}.

\bibitem[{\citenamefont{Kribs et~al.}(2010)\citenamefont{Kribs, Roy, Terning,
  and Zurek}}]{Kribs:2009fy}
\bibinfo{author}{\bibfnamefont{G.~D.} \bibnamefont{Kribs}},
  \bibinfo{author}{\bibfnamefont{T.~S.} \bibnamefont{Roy}},
  \bibinfo{author}{\bibfnamefont{J.}~\bibnamefont{Terning}}, \bibnamefont{and}
  \bibinfo{author}{\bibfnamefont{K.~M.} \bibnamefont{Zurek}},
  \bibinfo{journal}{Phys. Rev.} \textbf{\bibinfo{volume}{D81}},
  \bibinfo{pages}{095001} (\bibinfo{year}{2010}), \eprint{0909.2034}.

\bibitem[{\citenamefont{Forestell et~al.}(2017)\citenamefont{Forestell,
  Morrissey, and Sigurdson}}]{Forestell:2016qhc}
\bibinfo{author}{\bibfnamefont{L.}~\bibnamefont{Forestell}},
  \bibinfo{author}{\bibfnamefont{D.~E.} \bibnamefont{Morrissey}},
  \bibnamefont{and}
  \bibinfo{author}{\bibfnamefont{K.}~\bibnamefont{Sigurdson}},
  \bibinfo{journal}{Phys. Rev.} \textbf{\bibinfo{volume}{D95}},
  \bibinfo{pages}{015032} (\bibinfo{year}{2017}), \eprint{1605.08048}.

\bibitem[{\citenamefont{Witten}(1984)}]{Witten:1984rs}
\bibinfo{author}{\bibfnamefont{E.}~\bibnamefont{Witten}},
  \bibinfo{journal}{Phys. Rev.} \textbf{\bibinfo{volume}{D30}},
  \bibinfo{pages}{272} (\bibinfo{year}{1984}).

\bibitem[{\citenamefont{Lucini et~al.}(2004)\citenamefont{Lucini, Teper, and
  Wenger}}]{Lucini:2003zr}
\bibinfo{author}{\bibfnamefont{B.}~\bibnamefont{Lucini}},
  \bibinfo{author}{\bibfnamefont{M.}~\bibnamefont{Teper}}, \bibnamefont{and}
  \bibinfo{author}{\bibfnamefont{U.}~\bibnamefont{Wenger}},
  \bibinfo{journal}{JHEP} \textbf{\bibinfo{volume}{01}}, \bibinfo{pages}{061}
  (\bibinfo{year}{2004}), \eprint{hep-lat/0307017}.

\bibitem[{\citenamefont{Lucini et~al.}(2005)\citenamefont{Lucini, Teper, and
  Wenger}}]{Lucini:2005vg}
\bibinfo{author}{\bibfnamefont{B.}~\bibnamefont{Lucini}},
  \bibinfo{author}{\bibfnamefont{M.}~\bibnamefont{Teper}}, \bibnamefont{and}
  \bibinfo{author}{\bibfnamefont{U.}~\bibnamefont{Wenger}},
  \bibinfo{journal}{JHEP} \textbf{\bibinfo{volume}{02}}, \bibinfo{pages}{033}
  (\bibinfo{year}{2005}), \eprint{hep-lat/0502003}.

\bibitem[{\citenamefont{Megevand and Ramirez}(2017)}]{Megevand:2016lpr}
\bibinfo{author}{\bibfnamefont{A.}~\bibnamefont{Megevand}} \bibnamefont{and}
  \bibinfo{author}{\bibfnamefont{S.}~\bibnamefont{Ramirez}},
  \bibinfo{journal}{Nucl. Phys.} \textbf{\bibinfo{volume}{B919}},
  \bibinfo{pages}{74} (\bibinfo{year}{2017}), \eprint{1611.05853}.

\bibitem[{\citenamefont{Lesgourgues and Pastor}(2006)}]{Lesgourgues:2006nd}
\bibinfo{author}{\bibfnamefont{J.}~\bibnamefont{Lesgourgues}} \bibnamefont{and}
  \bibinfo{author}{\bibfnamefont{S.}~\bibnamefont{Pastor}},
  \bibinfo{journal}{Phys. Rept.} \textbf{\bibinfo{volume}{429}},
  \bibinfo{pages}{307} (\bibinfo{year}{2006}), \eprint{astro-ph/0603494}.

\bibitem[{\citenamefont{Pitaevskii and Lifshitz}(1981)}]{Pitaevskii}
\bibinfo{author}{\bibfnamefont{L.~P.} \bibnamefont{Pitaevskii}}
  \bibnamefont{and} \bibinfo{author}{\bibfnamefont{E.~M.}
  \bibnamefont{Lifshitz}}, \emph{\bibinfo{title}{{COURSE OF THEORETICAL
  PHYSICS: PHYSICAL KINETICS}}} (\bibinfo{publisher}{Pergamon Press},
  \bibinfo{address}{Oxford, U.K.}, \bibinfo{year}{1981}), ISBN
  \bibinfo{isbn}{0080206417}.

\bibitem[{\citenamefont{Cyr-Racine et~al.}(2016)\citenamefont{Cyr-Racine,
  Sigurdson, Zavala, Bringmann, Vogelsberger, and
  Pfrommer}}]{Cyr-Racine:2015ihg}
\bibinfo{author}{\bibfnamefont{F.-Y.} \bibnamefont{Cyr-Racine}},
  \bibinfo{author}{\bibfnamefont{K.}~\bibnamefont{Sigurdson}},
  \bibinfo{author}{\bibfnamefont{J.}~\bibnamefont{Zavala}},
  \bibinfo{author}{\bibfnamefont{T.}~\bibnamefont{Bringmann}},
  \bibinfo{author}{\bibfnamefont{M.}~\bibnamefont{Vogelsberger}},
  \bibnamefont{and} \bibinfo{author}{\bibfnamefont{C.}~\bibnamefont{Pfrommer}},
  \bibinfo{journal}{Phys. Rev.} \textbf{\bibinfo{volume}{D93}},
  \bibinfo{pages}{123527} (\bibinfo{year}{2016}), \eprint{1512.05344}.

\end{thebibliography}
\end{document}